\documentclass[12pt,preprint]{aastex}

\slugcomment{}

\shorttitle{new HST/ACS observations of NGC~4449}
\shortauthors{Annibali et al.}

\begin{document}


\title{Starbursts in the Local Universe: 
new HST/ACS observations of the irregular galaxy NGC~4449\footnote{Based on observations 
with the NASA/ESA {\it Hubble Space Telescope}, obtained at the 
Space Telescope Science Institute, which is operated  by AURA, Inc.,
for NASA under contract NAS5-26555.}}


\author{F. Annibali \altaffilmark{2}, A. Aloisi \altaffilmark{2,3}, 
J. Mack\altaffilmark{2}, M. Tosi \altaffilmark{4}, R.P. van der Marel\altaffilmark{2}, 
L. Angeretti\altaffilmark{4}, C. Leitherer\altaffilmark{2}, M. Sirianni\altaffilmark{2,3}} 

\altaffiltext{2}{Space Telescope Science Institute, 3700 San Martin Drive, 
Baltimore, MD 21218, USA; annibali@stsci.edu}

\altaffiltext{3}{On assignment from the Space Telescope Division
of the European Space Agency}

\altaffiltext{4}{INAF-Osservatorio Astronomico di Bologna, 
Via Ranzani 1, I-40127 Bologna, Italy}



\begin{abstract}

We present photometry 
with the Advanced Camera for Surveys (ACS)
on the Hubble Space Telescope (HST) 
of stars in the
Magellanic starburst galaxy NGC~4449. The galaxy has been imaged in
the F435W (B), F555W (V) and F814W (I) broad-band filters, and in the
F658N (H$\alpha$) narrow-band filter.
Our photometry includes $\approx$ 300,000 objects in the (B, V) 
color-magnitude diagram (CMD)
down to V $\la$ 28, and $\approx$ 400,000 objects in the (V, I) CMD, 
down to I $\la$ 27 . A subsample of $\approx$ 200,000 stars has been 
photometrized in all the three bands simultaneously.
The features observed in the CMDs imply a variety of stellar 
ages up to at least 1 Gyr, and possibly as old as a Hubble time.
The spatial variation of the CMD morphology
and of the red giant branch colors 
point toward the presence of an age
gradient: young and intermediate-age stars tend to be 
concentrated toward the
galactic center, while old stars are present everywhere.
The spatial variation in the average luminosity of 
carbon stars suggests that there is not a strong metallicity
gradient ($\lesssim 0.2$ dex). 
Also, we detect an interesting resolved star cluster 
on the West side of the galaxy, surrounded by a symmetric
tidal or spiral feature consisting of young stars.
The positions of the stars in NGC~4449 younger than 10 Myr  
are strongly correlated with the H$\alpha$ emission.
We derive the distance of NGC~4449 
from the tip of the red giant branch to be ${\rm D=3.82 \pm 0.27}$ Mpc. 
This result is in agreement with the distance
that we derive from the luminosity of the carbon stars.
\end{abstract}


\keywords{galaxies: dwarf --- galaxies: individual (NGC~4449) 
---galaxies: irregular ---galaxies: resolved stellar populations 
---galaxies: starburst }



\section{Introduction}

Starbursts are short and intense episodes of star
formation (SF) that usually occur in the central regions of galaxies
and dominate their integrated light.  The associated star-formation
rates (SFR) are so high that the existing gas supply can sustain the
stellar production only on timescales much shorter than a cosmic time
($\lesssim 1$ Gyr). 

The importance of the starburst phenomenon in the context of
cosmology and galaxy evolution has been
dramatically boosted in recent years by 
deep imaging and spectroscopic surveys 
which have discovered star-forming galaxies at high redshift:
a population of dusty and massive 
starbursts, with SFRs as
high as $\sim$ 100 -- 1000  M$_{\odot}$ yr$^{-1}$, 
has been unveiled in the submillimeter and millimeter 
wavelengths at z$>$2 \citep{blain02,scott02}
and star-forming galaxies at $z >$ 3 
have been discovered with the Lyman break selection technique 
\citep{steidel96,pet01} and through Lyman-$\alpha$
emission surveys (\citet{rhoads},  
see also \citet{lefevre05} for a more recent independent approach).

In the local Universe, starbursts are 
mostly found in dwarf irregular galaxies, 
and contribute $\sim$ 25\% of the whole massive SF \citep{heck98}.
Both observations and theoretical models \citep{larson78,genz98,ni86}
show that strong starbursts 
are usually triggered by processes such as interaction 
or merging of galaxies, or by accretion of gas,
which probably played an important role
in the formation and evolution of galaxies at high redshift.
Thus, nearby starbursts can serve as local
analogs to primeval galaxies to
test our ideas about SF,
evolution of massive stars, and physics of the interstellar medium (ISM)
in "extreme" environments.
The high spatial resolution and high sensitivity of 
Hubble Space Telescope offer the possibility 
to study the evolution of nearby starbursts
in details. This is fundamental in order to address many  
of the still open questions in cosmological astrophysics:
What are the main characteristics of primeval galaxies? 
What is the nature of star-forming galaxies at high redshift?
How important are accretion and  merging processes
in the formation and evolution of galaxies?

The Magellanic irregular galaxy NGC~4449
($\alpha_{2000} =Ê12^h 28^m 11^{s}.9$, 
$\delta_{2000} =Ê+ 44^{\circ} 05^{'} 40^{"}$, $l=136.84$ 
and $b=72.4$), 
at a distance of $3.82 \pm 0.27$ Mpc
(see Section~5), is one of the best studied and spectacular
nearby starbursts. 
It has been observed across the whole electromagnetic spectrum 
and displays
both interesting and uncommon properties. It is one of the most
luminous and active irregular galaxies. Its integrated magnitude $M_B
= -18.2$ makes it $\approx$ 1.4 times as luminous as the 
Large Magellanic Cloud (LMC) \citep{hunter97}.
\citet{th87} estimated a current SFR of 
$\sim 1.5$ M$_{\odot}$ yr$^{-1}$.
NGC~4449 is also
the only local example of a global starburst, in the sense that the
current SF is occurring throughout the galaxy \citep{hunter97}. 
This makes NGC~4449 more similar to Lyman break Galaxies (LBGs)  
at high redshift ($z \simeq3$), where the brightest regions of 
SF are embedded in a more diffuse
nebulosity and dominate the integrated light also at optical
wavelengths \citep{gi02}.

Abundance estimates in NGC~4449 
were derived in the HII regions by \cite{talent},  
\cite{hgr82} and \cite{mar97}, and for 
NGC~4449 nucleus by \citet{bok01}.
The published values are in good agreement with 
each other, and provide  12 + log(O/H) $\approx 8.31$. 
Adopting the oxygen solar abundance from \citet{sun98}, 
12 + log(O/H)$_{\odot} =$ 8.83,
we obtain [O/H] $=$ -0.52, 
i.e. NGC~4449 oxygen content 
is almost one third solar, as in the LMC.
New solar abundance estimates, based on 3D hydrodynamic
models of the solar atmosphere, accounting
for departures from LTE, and on improved atomic and molecular data,
provide 12 + log(O/H)$_{\odot} =$ 8.66 \citep{sun07}.
However, the new lower abundances seem to be inconsistent with 
helioseismology data, unless the majority
of the inputs needed to make the solar model are changed \citep{basu07}.
Thus, we will adopt the old abundances from \citet{sun98}
throughout the paper.

Radio observations of NGC~4449 have shown a very extended HI 
halo ($\sim 90$ kpc in diameter) which is a factor of $\sim 10$ 
larger than the optical diameter of the galaxy and appears to 
rotate in the opposite direction to the gas in 
the center \citep{baj94}. 
Hunter et al.~(1998, 1999) have resolved this halo 
into a central disk-like feature and large gas streamers that 
wrap around the galaxy. Both the morphology and the dynamics of 
the HI gas suggest that NGC~4449 has undergone some interaction 
in the past. A gas-rich companion galaxy, DDO~125, at the projected 
distance of $\sim 40$ kpc, could have been involved \citep{theis}.

NGC~4449 has numerous ($\sim 60$) star clusters  
with ages up to 1 Gyr \citep{gel01} and a young ($\sim$ 6-10 
Myr) central cluster \citep{bok01}, a prominent stellar 
bar which covers a large fraction of the optical body 
\citep{hun99}, and a spherical distribution of older (3-5 Gyr) stars 
\citep{both96}. The galaxy has also been demonstrated to contain 
molecular clouds from CO observations \citep{ht96} 
and to have an infrared (10-150 $\micron$) luminosity of $2 \times 
10^{43}$ erg s$^{-1}$ \citep{th87}. The ionized gas shows 
a very turbulent morphology with filaments, shells and bubbles which 
extend for several kpc (Hunter \& Gallagher 1990, 1997). The kinematics 
of the HII regions within the galaxy is chaotic, again suggesting 
the possibility of a collision or merger \citep{va02}.
Some 40\% of the X-ray emission in NGC 4449 comes from hot
gas with a complex morphology similar to that observed 
in H$\alpha$, implying an expanding super-bubble with a velocity 
of $\sim 220$ km\,s$^{-1}$ \citep{sum03}.

All these observational data suggest that 
the late-type galaxy NGC~4449 
may be changing as a result of an 
external perturbation, i.e., interaction or merger with another 
galaxy, or accretion of a gas cloud.
A detailed study of the star-formation history 
(SFH) of this galaxy is fundamental in order to derive a 
coherent picture for its evolution, and understand the 
connection between possible merging/accretion processes
and the global starburst.
With the aim of inferring its SFH,
we have observed NGC~4449 with the 
Advanced Camera for Surveys (ACS)
on the Hubble Space Telescope (HST) in the F435W, F555W, 
F814W and F658N filters. In this paper we present the new data
and the resulting color--magnitude diagrams (CMDs) (Sections~2, 3 and 4).
We derive a new estimate of the distance modulus 
from the magnitude of the tip of the red giant branch (TRGB) 
and the average magnitude of the carbon stars 
in Section~5. In Section~6, the empirical CMDs are compared
with stellar evolutionary tracks. With the use of the tracks,
we are able to derive the spatial distribution of stars
of different age in the field of NGC~4449.
The Conclusions are presented in Section~7. 
The detailed SFH of NGC~4449 will be derived through 
synthetic CMDs in a forthcoming paper.

\section{Observations and data reduction}

The observations were performed in November 2005 with
the ACS Wide Field Camera (WFC) using the F435W (B), F555W (V) and F814W (I) 
broad-band filters, and the F658N (H$\alpha$)
narrow-band filter (GO program 10585, PI Aloisi).
We had two different pointings
in a rectangular shape along the major axis of the galaxy.
Each pointing was organized with a 4 - exposure 
half $+$ integer pixel dither pattern
with the following offsets in arcseconds:
(0,0) for exposure 1, (0.12, 0.08) for exposure 2,
(0.25, 2.98) for exposure 3, and (0.37, 3.07)
for exposure 4. This dither pattern is suitable to remove
cosmic rays and hot/bad pixels, fill the gap between the
two CCDs of the ACS/WFC, and improve the PSF sampling.
The exposure times in the different broad-band filters
were chosen to reach at least one magnitude fainter than the TRGB, 
at I $\approx $24, with a photometric error below 0.1 mag (S/N $>$ 10).
Eight exposures of $\sim$ 900 s, 600 s, 500 s and
90 s were acquired for each of the B, V, I and 
H$\alpha$ filters, respectively.

For each filter, the eight dithered frames, calibrated
through the most up-to-date version of the 
ACS calibration pipeline (CALACS), 
were co-added into a single mosaicked image using the
software package MULTIDRIZZLE \citep{Koe02}.
During the image combination, we 
fine-tuned the image alignment, accounting 
for shifts, rotations, and scale variations between images. 
The MULTIDRIZZLE procedure also corrects
the ACS images for geometric distortion and
provides removal of cosmic rays and bad pixels.
The total field of view of the resampled 
mosaicked image is $\sim$ 380 $\times$ 200 arcsec$^2$, 
with a pixel size of 0.035~\arcsec (0.7 times the 
original ACS/WFC pixel size). 

To choose the optimal drizzle parameters,
we experimented with different combinations of the 
MULTIDRIZZLE parameters {\it pixfrac} (the linear size of the ``drop" in the input pixels)
and {\it pixscale} (the size of output pixels).
One must choose a {\it pixfrac} value that is small enough to avoid 
degrading the final image, but large enough that, when all images are 
dropped on the final frame, the flux coverage of the output image is fairly uniform. 
Statistics performed on the final drizzled weight image should 
yield an rms value which is less than 20\% of the median value.
In general, the {\it pixfrac} should be slightly larger than the scale 
value to allow some of the 'drop' to spill over to adjacent pixels. 
Following these guidelines, we find that {\it pixfrac}$=$0.8 and {\it 
pixscale}$=$0.7 provide the best resolution and PSF sampling for our dithered images.

\begin{figure}
\epsscale{1.}
\plotone{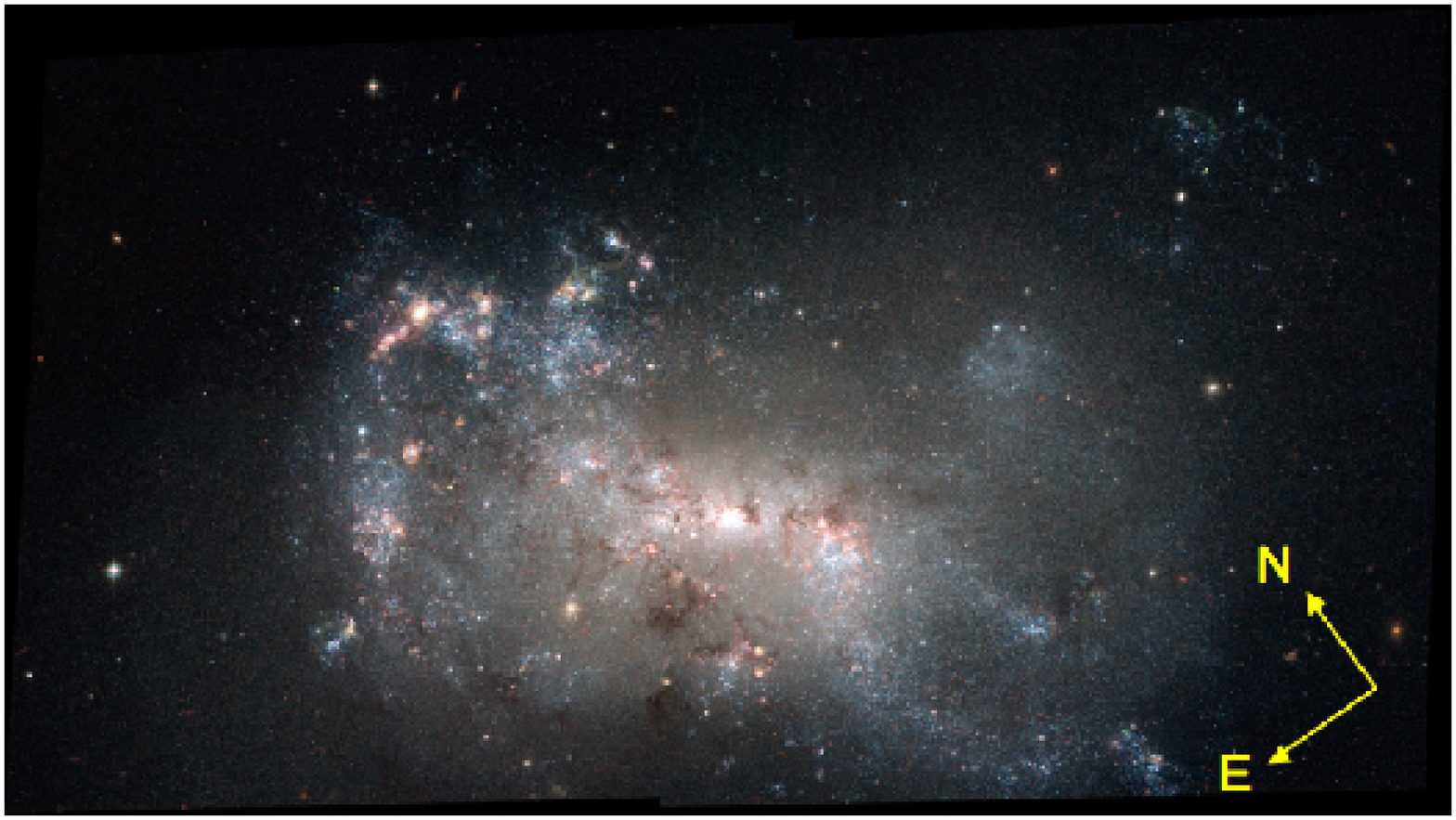}
\caption{Four-color 
(F435W (B), F555W (V), F814W (I) and F658N (H$\alpha$)) 
composite image of NGC~4449 showing a field of view of 
$\sim$ 380 $\times$ 200 arcsec$^2$, obtained from the mosaic
of the two different ACS pointings.
the B, V, I and H$\alpha$ 
data are shown in blue, green,
red and pink, respectively. The image orientation
is also indicated.
\label{image}}
\end{figure}

The total integration times are $\sim$ 3600 s, 2400 s, 
2000 s and 360 s for the B, V, I and H$\alpha$ images, respectively.
Only in a small region of overlap between the two pointings
($\sim$ 30 $\times$ 200 arcsec$^2$) the integration times are
twice as those listed above.
Figure~\ref{image} shows the mosaicked true-color image created by combining the
data in the four filters.

The photometric reduction of the images was performed with the
DAOPHOT package \citep{daophot} in the IRAF environment\footnote{IRAF is distributed
by the National Optical Astronomy Observatories, which are operated by
AURA, Inc., under cooperative agreement with the National Science 
Foundation}. 
The instrumental magnitudes were estimated via a 
PSF-fitting technique. We constructed 
a PSF template for each of the four ACS chips
(2 CCDs at 2 pointings) contributing to the final mosaicked image.
To derive the PSF, we selected
the most isolated and clean stars, 
uniformly distributed within each chip.
The PSF is modeled with an analytic moffat function 
plus additive corrections derived from
the residuals of the fit to the PSF stars.
The additive corrections include the first and
second order derivatives of the PSF with respect to 
the X and Y positions in the image. 
This procedure allows us to properly model
the spatial variation of the PSF in the ACS/WFC
field of view.

\begin{figure}
\epsscale{.80}
\plotone{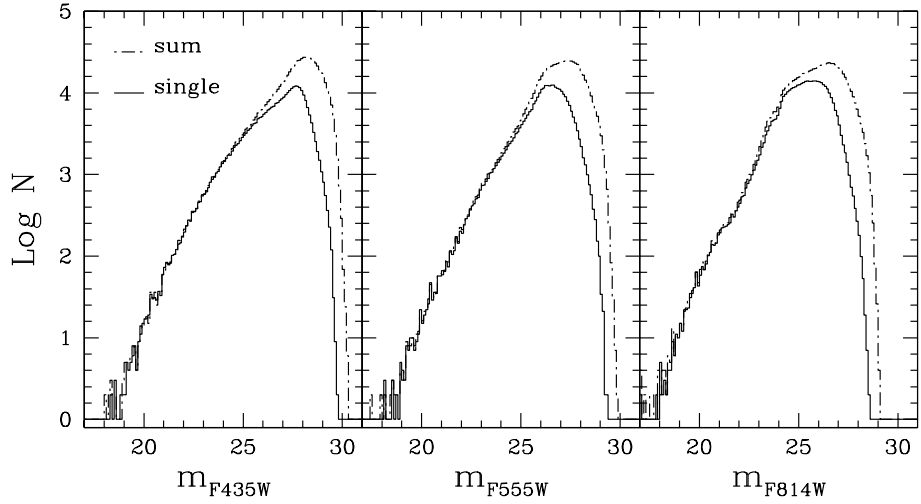}
\caption{Luminosity functions (LFs) in the 
F435W, F555W, and F814W filters. The solid line
is the LF obtained if the star detection is performed 
independently on the three different images.
The dot-dashed line refers to a deeper photometry,
obtained by detecting the stars in the sum of the 
F435W, F555W, and F814W images. 
\label{lfs}}
\end{figure}

The stars were detected independently
in the three bands, without forcing in the
shallowest frames the detection of the objects  
found in the deepest one. 
For comparison, we also ran the photometry 
on a list of stars detected on the sum of the  B, V and I images.
The luminosity functions (LFs) obtained in the 
various bands with the two different approaches 
are presented in Fig.~\ref{lfs}. 
We notice that the {\it forced} search
pushes the detection of stars $\sim$ 0.5 magnitude deeper
than the {\it independent} search.
On the other hand, upon closer inspection the majority of the ``gained" objects
turn out to be spurious detections or stars
with large photometric errors.
Furthermore, the deeper photometry is not deep enough
to detect the next features of interest in the CMD 
(the horizontal branch, the red clump or the 
asymptotic giant branch (AGB) bump, 
see Section~3), 
and thus it does not provide any additional information
for our study. In the following we thus use the photometry
obtained with the independent search on the three bands.

Aperture photometry with PHOT, and then
PSF-fitting photometry with the ALLSTAR package,
were performed at the position of the objects detected
in the B, V and I images.
The instrumental magnitudes were measured adopting
the appropriate PSF model to fit the stars 
according to their position in the frame.
The B,V and I catalogs were then cross-correlated
with the requirement of a spatial offset
smaller than 1 pixel between the positions of the stars
in the different frames.
This led to 299,115 objects having a measured magnitude 
in both B and V, 402,136 objects in V and I, and 213,187
objects photometrized in all the three bands simultaneously.

The conversion  of the instrumental magnitudes 
$m_i$ to the HST VEGAMAG system was performed by following
the prescriptions in \citet{sir05}. The HST VEGAMAG magnitudes 
are derived according to the equation:

\begin{equation}
m = m_i + C_{ap} + C_{\inf} + ZP + C_{CTE},
\label{eq1}
\end{equation}
where $m_i$ is the DAOPHOT magnitude 
(${\rm -2.5 \times \log(counts/exptime)}$) within a circular aperture 
of 2--pixel radius, and with the sky value computed in an annulus 
from 8 to 10 pixels; $C_{ap}$ is the aperture correction
to convert the photometry from the 2 pixel to the conventional $0.5\arcsec$
\ radius aperture, and with the sky computed at ``infinite"; 
we computed $C_{ap}$ from isolated stars selected in our images;
$C_{\inf}$ is taken
from \citet{sir05} and is an offset to convert the magnitude from the 0.5 \arcsec radius into a 
nominal infinite aperture; ZP is the HST VEGAMAG zeropoint for the given filter. 
Corrections for imperfect charge transfer efficiency (CTE)
were calculated from each single expousure, and then averaged,
following the formulation of \cite{cte}
(Eq. \ref{eq2}), which accounts for the time dependence of the 
photometric losses:

\begin{equation}
C_{CTE}=10^A \times SKY^B \times FLUX^C \times \frac{Y}{2048} \times \frac{MJD - 52333}{365},
\label{eq2}
\end{equation}
where SKY is the sky counts per pixel per exposure, FLUX is the star
counts per exposure within our adopted photometry aperture
(r$=$2 pixel in the resampled drizzled image, corresponding to 1.4 pixel
in the original scale),
Y is the number of charge transfers, and MJD is the Modified Julian Date.
The coefficients of equation~(\ref{eq2}) 
were extrapolated from a r$=$3 pixel aperture to a
r$=$1.4 pixel aperture (in the original scale), and 
are $A=1.08$, $B=-0.309$, and $C =-0.976$.
The computed CTE corrections are negligible for the brightest stars,
but can be as high as $\sim$ 0.1 mag for the faintest stars.
We did not transform the final magnitudes to the Johnson-Cousins
B, V, I system, since this would introduce additional uncertainties.
However, such transformations can be done in straightforward manner
using the prescriptions of \citet{sir05}.
The ACS VEGAMAG magnitudes were not corrected for 
Galactic foreground extinction 
(${\rm E(B-V) = 0.019}$, \citep{schlegel}) 
and internal reddening.
Concerning the internal reddening,
\citet{hill98} derived ${\rm E(B-V) \approx}$ 0.18
from the $H\alpha$/H$\beta$ ratio measured in NGC~4449 HII regions.
This value can be considered an upper limit to the average
internal extinction, since 
the nebular gas is usually associated with young star forming
regions, which tend to be inherently more dusty than 
the regions in which
older stars reside (as demonstrated explicitly for the case of
the LMC; \citet{zari}). 

\begin{figure}
\epsscale{.80}
\plotone{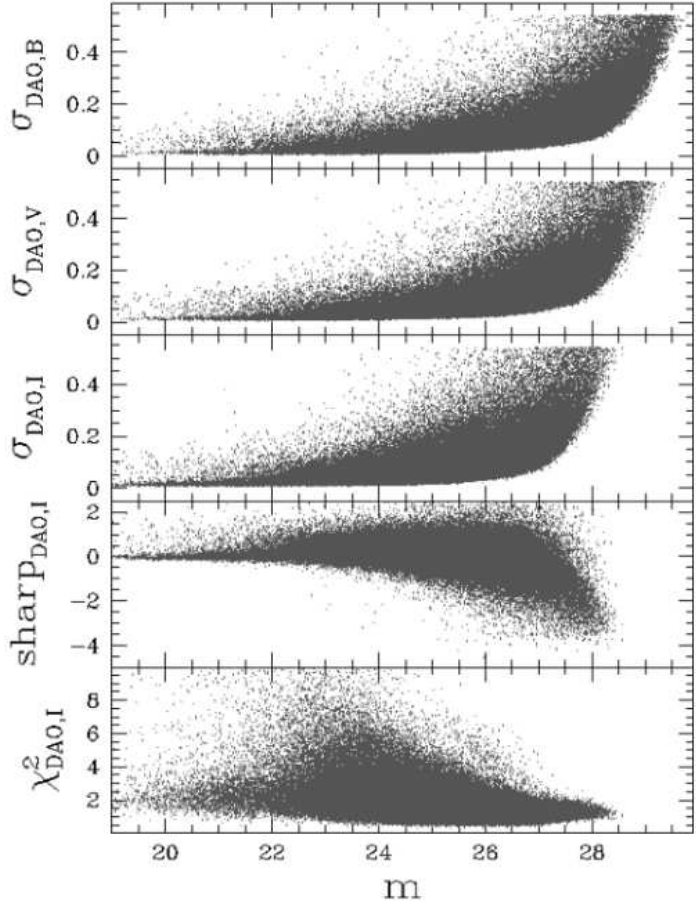}
\caption{Distribution of the DAOPHOT parameters 
$\sigma$, {\it sharpness} and $\chi^2$ as a function 
of the magnitude. The $\sigma$ distributions
for the F435W (B), F555W (V) and F814W (I) 
filters are shown in the three
upper panels. The {\it sharpness} and $\chi^2$ 
distributions are shown only for the F814W filter,
but look very similar in F435W and F555W.
\label{fig1I}}
\end{figure}

Because of both the large number of dithered exposures and 
the conservative approach of the independent search on the
three images, our catalog is essentially free of instrumental artifacts
such as cosmic rays or hot pixels. 
The distribution of the DAOPHOT parameters $\sigma$, $\chi^2$ 
and {\it sharpness} is shown in 
Fig.~\ref{fig1I}.
The $\sigma$ parameter measures the uncertainty on the
magnitude, the $\chi^2$ is the residual per degree of freedom 
of the PSF-fitting procedure, and the {\it sharpness} provides a measure
of the intrinsic size of the object with respect 
to the PSF. 
Notice that the DAOPHOT/ALLSTAR package automatically rejects the objects
with  $\sigma > 0.55$.
The obtained distributions for 
$\sigma$,  $\chi^2$  and {\it sharpness}
suggest that the vast majority of the detected sources
are stars in the galaxy with a small contamination
from stellar blends and background galaxies.
There are some objects with very bright 
magnitudes (${\rm m_{F814W} }\la 22$) and  
{\it sharpness} $>$0.5 in Fig.~\ref{fig1I}
(this is also observed in the F435W and F555W
filters, for which we do not show the $\chi^2$  and {\it sharpness}
distributions).
Their {\it sharpness} values imply that
they have a larger intrinsic size than
the PSF, and thus may not be individual stars.
By visually inspecting these objects in all the images,
we recognized several candidate star clusters
and background galaxies. Some of the candidate
star clusters look like fairly round but extended objects;
some others present a central core, 
and are partially resolved into individual stars in the outskirts.
We detect at least 42 clusters in our data,
some of which look like very massive globular clusters.
The candidate clusters and the galaxies 
that were identified by eye were rejected 
from the photometric catalog.
We are left with 299,014 objects in the  
(B, V) catalog, 402,045  objects in the (V, I) catalog, and 213,099
objects photometrized in all the three bands.
We experimented with many other cuts in 
$\sigma$,  $\chi^2$  and {\it sharpness}, but
none of them affected the global appearance
of the CMDs and the detected evolutionary features.
A detailed study of the cluster properties 
will be presented in a forthcoming paper 
(Aloisi et al., in preparation).

\section{Incompleteness and blending}

To evaluate the role of incompleteness and blending in our data,
we performed artificial star experiments on the drizzle-combined frames,
following the procedure 
described by \cite{tosi01}.
These tests serve to probe observational effects associated with the
data reduction process, such as the accuracy of the photometric measurements,
the crowding conditions, and the ability of the PSF-fitting procedure
in resolving partially overlapped sources.
We performed the tests using to the following procedure.
We divided the frames into grids of cells of chosen width
(50 pixels) and randomly added one artificial star per cell at each run.
This procedure prevents the artificial stars 
to interfere with each other, and avoids to bias the experiments 
towards an artificial crowding not really present in the original frames.
The position of the grid is randomly changed at each run,
and after a large number of experiments the stars are
uniformly distributed over the frame.
In each filter, we assign to the artificial star a random input magnitude
between $m_1$ and $m_2$, with $m_1$ $\approx$ 3 mag brighter than the
brightest star in the CMD, and $m_2$ $\approx$ 3 mag fainter then the faintest
star in the CMD.
At each run, the frame is re-reduced following exactly the same 
procedure as for the real data. The output photometric catalog is
cross-correlated with a sum of the original photometric catalog of real stars 
and the list of the artificial stars added into the frame.
This prevents cross-correlation of  
artificial stars in the input list with real stars recovered in the
output photometric catalog.
We simulated about half a million stars 
for each filter.
At each magnitude level, the completeness of our photometry 
is computed as the ratio of the number of recovered artificial stars
over the number of added ones. 
The completeness levels in the color magnitude diagrams (see Section~4)
are the product of the completeness factors in the two involved passbands.

\begin{figure}
\epsscale{1.}
\plotone{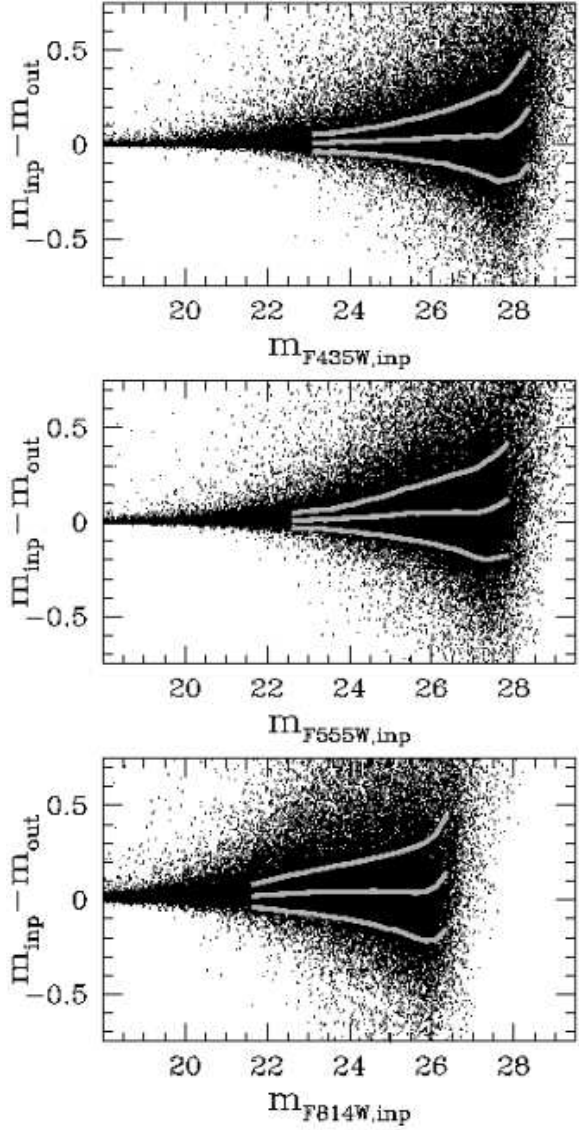}
\caption{Input minus output magnitude versus input
magnitude of the artificial stars in the F435W (top), 
F555W (middle) and F814W (bottom) filters. 
For each filter, we simulated half of a million stars.
The lines superimposed on the diagrams represent the mean 
$\Delta m$ and the $\pm 1$ standard deviations.
\label{dm}}
\end{figure}

We show in Fig.~\ref{dm} the $\Delta m$ difference between the input and
output magnitudes of the artificial stars as a function of
the input magnitude, for the F435W, F555W and F814W filters.
The solid lines superimposed on the artificial star distributions
correspond to the mean $\Delta m$ (central line), and the
$\pm 1 \sigma_m$ values around the mean.
The plotted $\Delta m$ distributions provide a complete and 
statistically robust characterization of the 
photometric error as a function of magnitude, for each filter. 
By comparing the $\sigma_m$ 
with the  DAOPHOT errors in Fig.~\ref{fig1I},
it is apparent that the DAOPHOT package 
increasingly underestimates the actual errors toward
fainter magnitudes. 
For instance, for a star with V$\sim$ 25.5 and I$\sim$ 24, 
(tip of the red giant branch, see Section~5), the mean DAOPHOT
error is $\sim$ 0.05 mag in both bands,
while the $\sigma_m$ from the artificial star tests is $\sim$ 0.15 
and $\sim$ 0.1 in V and I, respectively.
The systematic deviation from 0 of the mean 
$\Delta m$ indicates the increasing effect of blending, 
i.e. faint artificial stars recovered brighter than 
in input because they happen to overlap other faint objects.

\section{Color-magnitude diagrams}

\begin{figure}
\epsscale{1}
\plotone{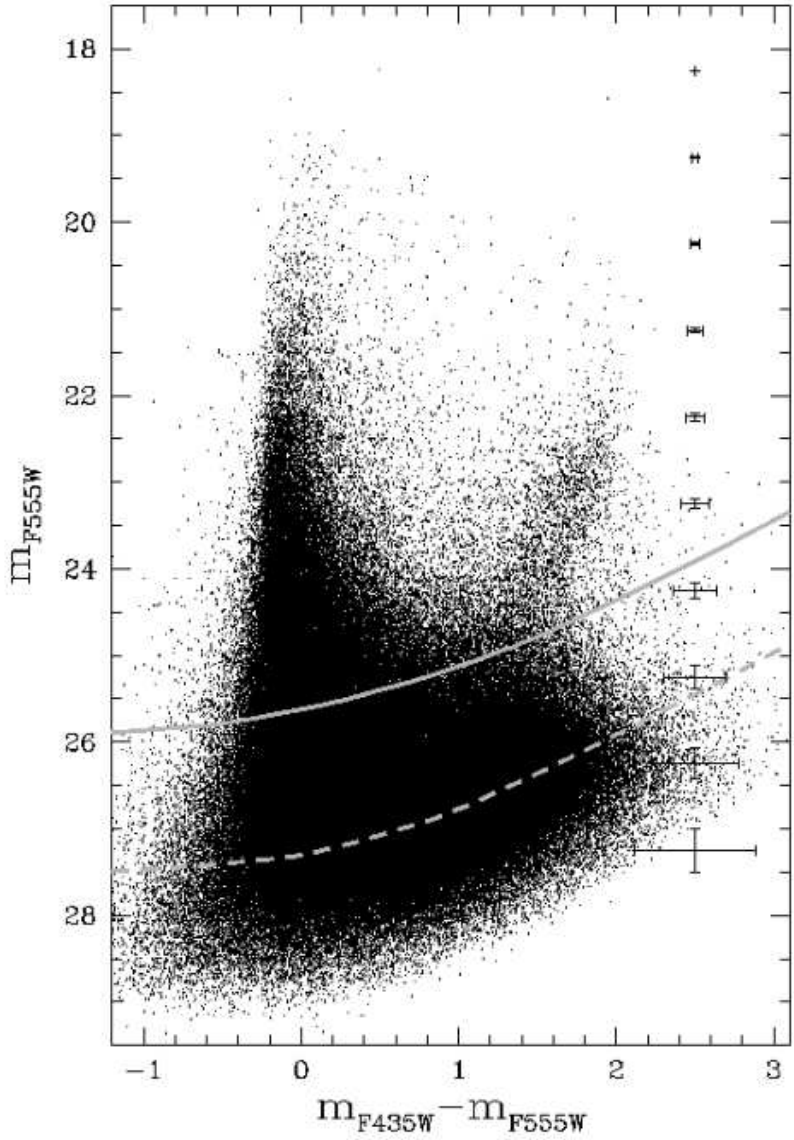}
\caption{${\rm m_{F555W}, m_{F435W}- m_{F555W}}$ CMD of the 
299,014 stars matched between the B and V catalogs.
The 90 \% (solid line) and 50 \% (dashed line)
completeness levels are indicated on the CMD.
The average size of the photometric errors, as derived
from the artificial star experiments, is indicated
at different ${\rm m_{F555W}}$ magnitudes.
The color error is evaluated at  ${\rm m_{F435W}- m_{F555W}=1}$.
\label{cmd1}}
\end{figure}

\begin{figure}
\epsscale{1}
\plotone{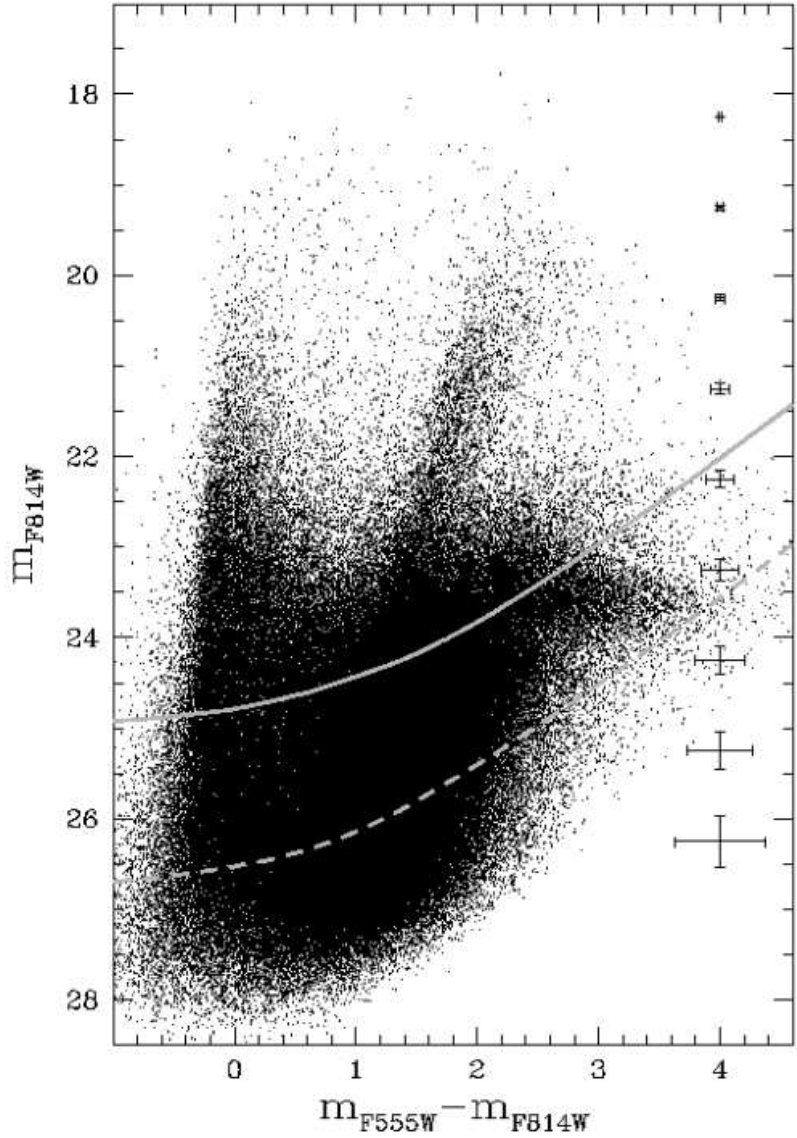}
\caption{${\rm m_{F814W}, m_{F555W}- m_{F814W}}$ CMD of the 
402,045 stars matched between the V and I catalogs.
The 90 \% (solid line) and 50 \% (dashed line)
completeness levels are indicated on the CMD.
The average size of the photometric errors, as derived
from the artificial star experiments, is indicated
at different ${\rm m_{F814W}}$ magnitudes.
The color error is evaluated at  ${\rm m_{F555W}- m_{F814W}=1}$.
\label{cmd2}}
\end{figure}

The CMDs are shown in Figures~\ref{cmd1} and \ref{cmd2}.
We plot the ${\rm m_{F555W}}$ versus 
${\rm m_{F435W}- m_{F555W}}$ CMD of the 
299,014 stars matched between the B and V catalogs in Fig.~\ref{cmd1},
and we plot the ${\rm m_{F814W}}$ versus ${\rm m_{F555W}-m_{F814W}}$ 
CMD of the 402,045 stars matched between the V and I catalogs
in  Fig.~\ref{cmd2}.
We indicate 
the 90 \% (solid line) and 50 \% (dashed line)
completeness levels as derived from the artificial star
experiments on the two CMDs.
The average size of the photometric errors
at different magnitudes,
as derived from artificial star tests,
is indicated as well.

The two CMDs show all the evolutionary features
expected at the magnitudes sampled by our data:
a well defined blue plume and red plume,
the red horizontal tail of the carbon stars in the
${\rm m_{F814W}}$ versus ${\rm m_{F555W}-m_{F814W}}$ CMD, and
a prominent red giant branch (RGB).
The blue plume is located at 
${\rm m_{F435W}-m_{F555W}}$ and ${\rm m_{F555W}-m_{F814W}}$  
$\simeq$ $-$0.1 in the two diagrams, with the brightest 
stars detected at ${\rm m_{F555W}}$, ${\rm m_{F814W}}$ $\sim$ 18. 
It samples both stars in the main-sequence (MS) evolutionary
phase and evolved stars at the hot edge of the core 
helium burning phase.
The blue plume extends down to the faintest magnitudes in our data,
at ${\rm m_{F555W} \sim 28}$.
The red plume is slightly inclined with respect 
to the blue plume, 
with {$\rm m_{F555W} \la 25$} and colors extending from 
{$\rm m_{F435W}-m_{F555W} \sim 1.4$} 
to $\sim$ 1.8 in the  
${\rm m_{F555W}}$, ${\rm m_{F435W}- m_{F555W}}$ CMD, 
and {$\rm m_{F814W} \la 23.5$} and colors 
extending from {$\rm m_{F555W}-m_{F814W} \sim 1.4$} to $\sim$ 2.2
in the ${\rm m_{F814W}}$ versus ${\rm m_{F555W}-m_{F814W}}$ CMD.
It is populated by 
red supergiants (RSGs) at the brighter magnitudes, 
and AGB stars at fainter
luminosities.
At intermediate colors, below ${\rm m_{F555W} \sim 25}$ and 
${\rm m_{F814W} \sim 23.5}$, we recognize 
the {\it blue loops} of intermediate-mass stars in the core 
helium burning phase.
The concentration of red stars at ${\rm m_{F555W} \ga 25.5}$
and ${\rm m_{F814W} \ga 24}$, corresponds to low-mass 
old stars in the RGB evolutionary phase. 
Finally, a pronounced horizontal feature, 
at ${\rm m_{F814W} \sim 23.5}$, and with colors extending  
from ${\rm m_{F555W}-m_{F814W} \sim 1.8}$ to as much as
${\rm m_{F555W}-m_{F814W} \sim 4}$, 
is observed in the ${\rm m_{F814W}}$, ${\rm m_{F555W}-m_{F814W}}$
CMD. This red tail is produced by carbon stars in the thermally pulsing 
asymptotic giant branch (TP-AGB) phase.

In order to reveal spatial differences in the stellar population 
of NGC~4449, we have divided the galaxy's field of view into 
28 (7 $\times$ 4) rectangular regions, as shown in Fig.~\ref{imagegrid}. 
The size of the regions ($\approx$ 55 $\times$ 55 ${\rm arcsec^2}$,
corresponding to ${\rm \approx 1 \times 1 \ kpc^2}$ 
at the distance of NGC~4449)
allows us to follow spatial variations at the kpc scale,
being at the same time large enough to provide a good sampling.
The ${\rm m_{F555W}}$, ${\rm m_{F435W}- m_{F555W}}$, and
${\rm m_{F814W}}$, ${\rm m_{F555W}- m_{F814W}}$ CMDs derived
for the different regions are shown in Fig.~\ref{hessregionbv} and 
\ref{hessregionvi}, respectively. 
The completeness levels plotted on the CMDs of the
central column show that 
the photometry is 
deeper in the external regions 
than in the galaxy center, where the high crowding level
makes the detection of faint objects more difficult.
The errors, as estimated from the artificial star
experiments, increase toward the galaxy center, as an effect
of the higher crowding level and 
the higher background.
Bright stars are mostly concentrated 
toward the galaxy center, and only a few of them are present 
at large galactocentric distances,
in agreement with what was already observed in other 
dwarf irregular galaxies, (e.g., \cite{tosi01}).
The external region  (6,4) 
makes the exception to this observed global trend, showing a
prominent blue plume in both the CMDs.
The luminous blue stars observed in these CMDs correspond in the image
of Fig.~\ref{image} to
a symmetric structure. This structure is more clearly visible in the
top right of Fig.~\ref{spatial}, which will be discussed 
in Section~6 below.
This structure could be due to tidal tails or spiral--like feature
associated with a dwarf galaxy that is currently being disrupted. The
structure is centered on a resolved cluster-like object that could be
the remnant nucleus of this galaxy. Fig.~\ref{imagegrid} 
shows a blow-up of this object.

\subsection{Carbon stars}

In the ${\rm m_{F814W}}$, ${\rm m_{F555W}- m_{F814W}}$ 
CMD of Fig.~\ref{cmd2},
the horizontal red tail, at magnitudes brighter than the 
TRGB, is due to carbon-rich stars in the TP-AGB phase. 
Since AGB stars trace the stellar populations 
fromÊ$\sim$ 0.1 to several Gyrs, this well defined feature 
is suitable to investigate the SFH from old to 
intermediate ages \citep{cioni06}. 
Recently, theoretical models of TP-AGB stars have been presented
by \cite{marigo07} for initial masses between 
0.5 and 5.0 $M_{\odot}$ and for different metallicities.
Their Fig.~20 shows that the position of the carbon--star
tracks in the $\log L/L_{\odot}$ vs $\log T_{eff}$ plane
depends on both mass and metallicity.
Higher-mass stars exhibit larger luminosities
for a given metallicity. On the other hand,
lower metallicities imply both a wider range of masses
undergoing the C-rich phase, and higher luminosities
for the more massive carbon stars.

From an empirical aspect, we can get the dependence of
the carbon--star luminosity on age and metallicity 
from the work of \cite{batti05} (hereafter BD05).
The authors provide the following relation
for the dependence of the carbon--star 
I band magnitude on metallicity:

\begin{equation}
{\rm <M_{I,carbon}>=-4.33 +0.28 \times [Fe/H]},
\label{eq3}
\end{equation}
which was derived through a least-square fit of the mean 
absolute I band magnitude of carbon stars in nearby galaxies with metallicities
${\rm -2<[Fe/H]<-0.5}$. According to (\ref{eq3}),
a drop of 1 dex in metallicity results
in a decrease of $\approx$ 0.3 
in the average magnitude of carbon stars.
The age dependence of the carbon--star luminosity is more
difficult to quantify since it requires a detailed knowledge 
of the SFH in galaxies. 
BD05 do not study such a dependence, but some qualitative 
considerations can be
obtained from examination of their Fig.~4.
The scatter of the ${\rm <M_{I,carbon}>}$ versus ${\rm [Fe/H]}$ relation 
is of the order of 0.1 mag (excluding  AndII and AndVII from the fit), 
i.e. $\approx$ 20 \% of the variation in  ${\rm <M_{I,carbon}>}$
spanned by the data.
We also notice that at fixed metallicity, galaxies with current star formation
(empty dots in Fig.~1 of BD05)
tend to have brighter ${\rm <M_{I,carbon}>}$ values, while galaxies
with no current star formation preferentially lie below the relation.
This suggests that at least part of the scatter 
of relation (\ref{eq3}) is due to a 
spread in age, with younger stellar populations having 
brighter carbon stars than older stellar populations.
This age dependence is consistent with 
the \cite{marigo07} models, where more massive
(younger) carbon stars tend to be more luminous.

Carbon stars were selected in NGC~4449 at 
${\rm 23< m_{F814W} <24, m_{555W}-m_{F814W}>2.4}$,
for each of the 28 regions shown in Fig.~\ref{imagegrid}.
The color limit was chosen to avoid a significant
contribution of RGB and oxygen-rich AGB stars.
For each region, we fitted the ${\rm m_{F814W}}$-band LF  
with a Gaussian curve, and adopted the peak of the 
best-fitting Gaussian as the average ${\rm <m_{I,carbon}>}$ 
in that region.
The errors on the results of the Gaussian fits
can be approximated as $\Delta m \approx \sigma / \sqrt{N}$,
where $\sigma$ is the width
of the Gaussian and N is the number of stars.
We experimented with different cuts in color 
(up to ${\rm m_{555W}-m_{F814W} >2.8}$) and different binnings of the data,
and found that none of them significantly affects 
the final results of the analysis.
The results presented in Fig.~\ref{Clum}  were obtained by binning 
the stellar magnitudes in bins of 0.2 mag. 
The quantity ${\rm \Delta m_{I,C}}$ along the 
ordinate is the difference between the carbon--star magnitude
measured in a specific region, and the 
carbon--star magnitude averaged over the whole field of view of NGC~4449.
Along the abscissa is the X coordinate in pixels.
From top to bottom, the panels refer to regions of decreasing Y
coordinate. Fig.~\ref{Clum} shows that the 
observed carbon--star luminosity 
is reasonably constant over almost the whole galaxy, within the errors.
The only significant variation is observed in the central regions
($ 3000 \la X \la 8000$, panel c), where 
the carbon stars appear to be up to $\approx$ 0.25 mag brighter than the average value.

We performed Monte Carlo simulations  
to understand if the observed variation of the C star LF
is intrinsic, i.e. due to stellar population gradients present 
in NGC~4449 field, or if it is an effect of the
photometric error and completeness level variations across the field.
We adopted an estimated {\it intrinsic} LF for the C stars,
that we assumed to be constant all over the field.
Then we investigated how the {\it intrinsic} LF is transformed
into the {\it observed} one after completeness and photometric errors 
were applied at different galacto-centric distances.
For simplicity, we assumed that the 
intrinsic ${\rm m_{F814}}$ distribution of the C stars  
is Gaussian, with parameters
(${\rm m_{F814W,0}=23.64}$, $\sigma$ $\sim 0.3$) 
derived from the Gaussian fit to the observed 
LF in the most external regions ((1,1:4), (1:7,4), (7,1:4);
 where this notation will indicate the some of these regions).
We also assumed that the C stars are not homogeneously
distributed in 
color, but follow a power law in ${\rm m_{F555W} - m_{F814W}}$.
The power law's parameters were 
derived by fitting the observed
C star color distribution in the 
same external regions.
Our assumption that the 
observed outer LF is a good description
of the intrinsic one in those regions is reasonable,
since the most external ${\rm m_{F814}}$, ${\rm m_{F555W} -m_{F814W}}$ 
CMDs are more than 90 \% complete
at the average C star luminosity, and the errors
$\sigma_{F814W}$ are only a few hundreds of magnitudes there
(see Fig.~\ref{hessregionvi}).
Monte Carlo extractions of (${\rm m_{F814}}$, ${\rm m_{F555W} - m_{F814W}}$)
pairs were drawn from
the assumed magnitude and color distributions.
A completeness and a photometric error
were then applied to each extracted star, for both an external 
and an internal region of NGC~4449.
The effect of incompleteness and photometric
errors on the distribution is shown in 
the bottom panel of Fig.~\ref{Clf}.
In the top panel 
we show instead the observed 
LFs for an external region 
((1,1:4), (1:7,4), (7,1:4)), and for the most internal one (4,3),
which displays the largest shift of the distribution peak
with respect to the external region.
Our simulations
in the bottom panel 
show that 
the C star LF
is mostly unaffected by incompleteness and photometric errors
in the most external regions.
This result testifies that the observed 
LF is very close to the intrinsic one in the periphery,
and that we adopted a reasonable input distribution
for the Monte Carlo simulations.
In the most internal region, instead,
the LF is shifted toward brighter magnitudes by an
amount (${\rm \Delta m_{F814W} \approx 0.25}$) 
comparable to the shift
between the observed LFs.
The resulting internal distribution 
also has a larger width ($\sigma \approx 0.4$)
than the intrinsic one, due to the
larger photometric errors.

Our results show that the 
detected change in the C star brightness 
over NGC~4449 field can be largely 
attributed to differences in completeness
between the center and the most external regions.
Accounting for this effect, 
the average magnitude of the C stars is constant
within the errors ($\approx 0.05$ mags).
From the BD05 relation
\footnote{
BD05 use the Johnson-Cousins I-band.
This is very similar to ${\rm m_{F814W}}$ \citep{sir05},
and the small difference can be ignored
for the purpose of the differential 
argument presented here.}, 
a change in magnitude of 0.05
corresponds to $\Delta$[Fe/H] $\approx 0.2$.
We interpret this as an upper limit to the
metallicity variation over the field of
view. This is consistent with 
studies of metallicity gradients 
in other magellanic irregulars. For example, 
\cite{cole04} derive a metallicity gradient of $\approx$ 
$-$0.05 dex $\times$ kpc$^{-1}$ in the LMC by comparing the abundances
in the inner disk and in the outer disk/spheroid \citep{ols},
while \citet{gro06} find no metallicity 
gradients from spectroscopic studies of cluster stars in the LMC.

\subsection{RGB stars}

In the CMD of Fig.~\ref{cmd2}, the morphology of the RGB, at
${\rm m_{F814W}} \ga 24$ and ${\rm m_{F555W}- m_{F814W} \ga 1}$, 
is connected to the properties of the stellar content
older than $\sim$ 1 Gyr.
Despite the poorer time resolution 
with increasing look-back time, and the well
known age-metallicity degeneracy, some constraints 
on the properties of the old stellar population can be inferred
from an analysis of the RGB morphology.

We derived the average RGB color as a function
of ${\rm m_{F814W}}$ by selecting stars 
with ${\rm m_{F814W}} \ga 24$,
and then performing a Gaussian fit to the  
${\rm m_{F555W}-m_{F814W}}$ color distribution 
at different magnitude bins.
The peak of the Gaussian fit is as red as 
${\rm m_{F555W}-m_{F814W} \approx 1.7}$ at the RGB tip,
and it is ${\rm m_{F555W}-m_{F814W} \approx 1.45}$ at 
${\rm m_{F814W}=25}$, one magnitude below the tip.
As expected, the RGB
is significantly redder than in more metal-poor star-forming galaxies
that we have previously studied with HST/ACS 
\citep{alo05,alo07}.

In order to reveal the presence of age/metallicity
gradients in NGC~4449, we performed a spatial
analysis of the RGB morphology,
following the same procedure as in
Section~4.1.
For each of the 28 rectangular 
regions identified in Fig.~\ref{imagegrid},
we performed a Gaussian fit to the RGB 
${\rm m_{F555W}- m_{F814W}}$ color distribution
for bins of ${\rm \Delta  m_{F814W}=0.25}$.
Then we averaged for each region the colors derived in the
four brightest bins, at  
${\rm 24 \le m_{F555W}- m_{F814W} \le 25}$,
i.e. one magnitude below the TRGB.
The results of our analysis are presented in Fig.~\ref{deltargb}.
We plot along the ordinate 
the difference between the 
RGB color in each region 
and the average RGB color 
in the total field of view of NGC~4449;
along the abscissa is the X coordinate in pixels.
From top to bottom, the panels refer to regions of decreasing Y
coordinate. 
Fig.~\ref{deltargb} shows 
that the RGB is bluer in the center
than in the periphery of NGC~4449, with variations
up to $\approx$ 0.3 mag in the ${\rm m_{F555W}-m_{F814W}}$ color.

This effect is also shown in the top panel 
of Fig.~\ref{rgbcolor},
where we plotted the observed color distributions of stars 
with ${\rm 24 \le m_{F814} \le 25}$,
for an external region ((1,1:4), (1:7,4), (7,1:4)), 
and for an internal region (3:5,2) of NGC~4449.
The peaks of the Gaussian fits to the external and internal distributions
differ by $\approx$ 0.26 mag, and their errors are very small 
($\sigma / \sqrt{N} \approx$  0.001 mag)
due to the large number of stars in each distribution
($\approx 20,000$).
The Gaussian fit is broader for the internal
region ($\sigma \approx 0.4$) than for the external region
($\sigma \approx 0.2$).
Also, the central region has a much broader tail of stars towards blue colors,
due to the contamination from younger blue-loop and MS stars.
The contribution from the MS $+$ blue--loop stars at the hot
edge of the core He-burning phase is recognizable as 
a bump at V$-$I $\approx 0$.

As was done for the C stars in Section~4.1,
we performed Monte Carlo simulations 
to understand if the observed difference 
in the color distributions
is intrinsic, or if it can be
attributed to the larger crowding in the central
regions of NGC~4449.
We assumed an initial distribution,
and drew Monte Carlo extractions 
from it with application of photometric errors and 
incompleteness.
As a first guess for the initial distribution,
we adopted a Gaussian with parameters
${\rm m_{F555W}-m_{F814W}=1.65}$ and $\sigma=0.15$.
This has the same mean but somewhat 
smaller $\sigma$ than the observed 
color distribution in the most external regions.
The simulated color distributions 
for the external and internal regions
were generated by applying the photometric 
errors and 
the completeness levels derived 
from artificial star experiments
in the considered regions of NGC~4449.
The results are presented in the central panel 
of Fig.~\ref{rgbcolor}.
As observed, the width of the simulated distribution 
in the internal region is larger
than in the external one 
because of the higher photometric errors.
While the peak of the external 
simulated distribution
is the same as that of the initial
distribution,
the peak of the internal
distribution is blueshifted by
an amount of $\approx$ 0.06 mags.
However, this is much less than the
observed shift of $\approx$ 0.3 mags.
The simulations therefore show that the 
completeness and the photometric error
variations over NGC~4449 field of view  
can account only in part 
($\approx$ 20 \% ) for the
shift between the observed 
internal and external distributions.
Thus the observed shift
must be mainly due to an intrinsic variation 
of the stellar population properties.

As a test, we performed a new simulation
starting from a Gaussian with 
a peak as blue as 
${\rm m_{F555W} - m_{F814}=1.45}$ and
with $\sigma=0.15$.
The result for the internal region 
is shown in the bottom panel of
Fig.~\ref{rgbcolor}. Here we 
plot also the observed color distribution,
for comparison.
Both the peaks and the widths
of the simulated and observed 
distributions are in good agreement.
The only discrepancy is observed at the 
bluest color, where of course we do not reproduce
the tail of the MS and post-MS stars.
Our simulations suggest that 
the peak of the color distribution  
toward the center is intrinsically 
bluer than in the periphery.
Assuming that the bluer peak 
in the center is due to a bluer 
RGB, the possible interpretations are 
1) younger ages; 2) lower metallicities; or 
3) lower reddening.
If differential reddening is present within NGC~4449,
we expect the central regions to 
be more affected by dust extinction
than the periphery,
and this would cause an even redder
RGB in the center.
Lower metallicities in NGC~4449 center are also very 
unlikely, since abundance determinations in galaxies
show that metallicity tends to decrease from the center
outwards or to remain flat (see Section~4.1). 
Thus, a bluer RGB would most likely indicate a younger
population in the center of NGC~4449.
Alternatively, the bluer peak observed in the center
could be due to ``contamination" by intermediate-age blue loop stars at 
the red edge of their evolutionary phase. But this too would 
imply the presence of
a younger stellar population in the center than in the
periphery of NGC~4449.
More quantitative results on age/metallicity
in NGC~4449 gradients
will be derived through fitting of synthetic CMDs,
and presented in a forthcoming paper
(Annibali et al. 2008, in preparation).

\section{A new distance determination}

The magnitude of the TRGB can be used to determine
the distance of NGC 4449. The top panel of Fig.~\ref{trgb} shows the
I-band LF of those stars in our final catalog
that have V$-$I in the range $1.0$--$2.0$. Here V and I are
Johnson-Cousins magnitudes, obtained from our magnitudes in the ACS
filter system using the transformations of
\cite{sir05} and
applying a foreground extinction correction of ${\rm E(B-V) = 0.019}$
\citep{schlegel}.
The TRGB is visually identifiable as the steep increase towards fainter
magnitudes at ${\rm I \approx 24}$. At this magnitude, RGB stars start to
contribute with a LF that increases roughly as a power law towards
faint magnitudes. By contrast, the stars in the LF at brighter
magnitudes are exclusively red supergiants and AGB stars. The drop at
magnitudes fainter than ${\rm I \approx 25.5}$ is due to incompleteness.

To determine the TRGB magnitude we used the software and methodology
developed by one of us (R.P.v.d.M) and described in detail in 
\cite{cioni00}. A discontinuity produces a peak in all of the higher-order
derivatives of the LF. We use a so-called Savitzky-Golay filter on the
binned LF to obtain the first- and second-order derivatives.  These
are shown in the middle and bottom panel of Fig.~\ref{trgb},
respectively. Peaks are indeed visible at
the expected position of the TRGB. We fit these with Gaussians and
find that the first derivative has a peak at ${\rm I_1 = 24.09}$, 
while the second derivative has a peak at ${\rm I_2 = 23.88}$. 
The reason for the
difference between these magnitudes is the presence of photometric
errors and binning in the analysis. This smooths out the underlying
discontinuity. As shown in Fig.~A.1 of \cite{cioni00}, this
causes the first derivative to overestimate the magnitude of the TRGB,
and the second derivative to underestimate the magnitude of the TRGB.
These biases can be explicitly corrected for as in Fig.~A.2 of 
\cite{cioni00}, using a simple model for the true underlying LF and the
measured width of the Gaussian peaks in the first- and second-order
derivatives of the LF. After application of this correction we obtain
the final estimates ${\rm I_{TRGB,1} = 23.99}$ and 
${\rm I_{TRGB,2} =24.00}$ for the underlying TRGB magnitude, 
based on the first and second derivatives, respectively. 
The good agreement between these
independent estimates shows that the systematic errors in the method
are small, in agreement with \cite{cioni00} who adopted a
systematic error ${\rm \Delta I_{TRGB} = \pm 0.02}$.
The additional systematic error introduced by 
the uncertainties in photometric
zeropoints, transformations, and aperture corrections
\citep{sir05} is ${\rm \Delta I_{TRGB} = \pm 0.03}$.
The random error on ${\rm \Delta I_{TRGB}}$ is very small due to 
the large number of stars
detected in NGC 4449. It can be estimated using bootstrap techniques
to be ${\rm \Delta I_{TRGB} = \pm 0.01}$.

Our estimate of the TRGB magnitude, ${\rm I_{TRGB} = 24.00 \pm 0.01}$
(random) $\pm 0.04$ (systematic),
can be compared to the absolute
magnitude of the TRGB, which was calibrated as a function of
metallicity by, e.g., \cite{bella04}. Adopting ${\rm [M/H] =
-0.52}$ for NGC 4449 (based on the oxygen abundance given in Section~1)
their calibration (top panel of their Fig.~5) predicts ${\rm M_{I, TRGB} =
-3.91}$. Comparison of different studies suggests that the systematic
uncertainty in this prediction is $\sim 0.15$. 
This takes into account also the possibility that the RGB star metallicity 
is actually lower than that of the HII regions (see discussion in Section 6).
The implied distance
modulus for NGC 4449 is therefore ${\rm (m-M)_0 = 27.91 \pm 0.15}$,
where we have added all sources of uncertainty in quadrature. This corresponds
to ${\rm D = 3.82 \pm 0.27}$ Mpc.

An alternative method for estimating the distance of NGC 4449 is
through the average I-band magnitude of the carbon
stars. Fig.~\ref{agbdist} shows the I-band luminosity function of
those stars in our final catalog that have V$-$I in the range
$2.2$--$3.0$.  There is a well-defined peak, due to the horizontal
``finger'' of carbon stars seen in the CMD of Fig.~\ref{cmd2}. A
Gaussian fit to the peak yields ${\rm I_{carbon} = 23.59 \pm 0.01}$,
corrected for foreground extinction. 
Adopting ${\rm [Fe/H] = -0.52}$ for NGC~4449 
(based on the oxygen abundance given in Section~1, and assuming for simplicity 
that oxygen traces the iron content), the BD05 calibration 
in (\ref{eq3}) predicts $M_{\rm I,
carbon} =-4.48$.  The systematic uncertainty in this prediction is
difficult to quantify. This is because carbon--star magnitudes are not
well understood on the basis of stellar evolution theory (by contrast
to the TRGB), because the dependence on stellar age or star formation
history is poorly quantified, and because only a few empirical
studies exist. We therefore adopt an uncertainty of $\sim 0.2$ mag,
consistent with the discussion of Section~4.1
The implied distance
modulus for NGC~4449 is then ${\rm (m-M)_0 = 28.07 \pm 0.20}$, 
where we have added all sources of uncertainty in
quadrature. This corresponds to ${\rm D = 4.11 \pm 0.38}$ Mpc. 
This is in agreement with the TRGB result, given the uncertainties.

We can compare our result of ${\rm D=3.82 \pm 0.27}$ Mpc to the previous
estimate of NGC~4449 distance by \cite{ka03}, who
inferred ${\rm D=4.2 \pm 0.5}$ Mpc. This result was derived by applying the
TRGB-luminosity method to HST/WFPC2 data.  The extinction-corrected
magnitude at which they detected the TRGB is 
${\rm I_{TRGB} = 24.07 \pm 0.26}$, which is consistent with 
our somewhat lower value of ${\rm I_{TRGB} = 24.00 \pm 0.01}$ 
(random) $\pm 0.04$ (systematic). Their
somewhat larger distance is also due to the different value adopted
for the absolute magnitude of the TRGB. They adopt ${\rm M_{I, TRGB} =-4.05}$
(as appropriate for metal-poor systems), while we adopt 
${\rm M_{I, TRGB} = -3.91}$ from the more recent 
\cite{bella04} calibration (which takes into account the 
metallicity dependence of the TRGB luminosity).

\section{Comparison with models}

For a direct interpretation of the CMDs in terms
of the stellar evolutionary phases,
we have superimposed stellar evolutionary tracks
for different metallicities  on the ${\rm m_{F555W}}$, ${\rm m_{F435W}- m_{F555W}}$, and
${\rm m_{F814W}}$, ${\rm m_{F555W}- m_{F814W}}$ CMDs 
(Figs.~\ref{cmdtracks} and~\ref{rgbtracks}).
The tracks at Z$=$0.008, Z$=$0.004, and Z$=$0.0004
are the Padua stellar evolutionary tracks 
(Fagotto et al. 1994a, 1994b) transformed into the
ACS Vegamag system by
applying the \citet{origlia} code,
and corrected for Galactic extinction (${\rm E(B-V)=0.019}$, 
\citet{schlegel}) 
and distance modulus (${\rm (m-M)_0 = 27.91}$, see Section~5).
The Z$=$0.001 tracks were obtained from the Padua tracks
through interpolation in metallicity \citep{ang06}.

The metallicity range covered by the plotted tracks
can account for the different populations potentially present in NGC~4449.
The Z$=$0.008 and Z$=$0.004 tracks  
are suitable to account for the more metal rich population,
given that the
abundances derived in the nucleus and disk of NGC~4449 are almost
one third of the solar value (see Section~1).
The plotted tracks are for masses in the range 0.9--40 ${\rm M_{\odot}}$.
Lower mass stars would not have the time to reach visible phases
within a Hubble time at the distance of NGC4449 (in these sets
a 0.8 ${\rm M_{\odot}}$ star reaches the TRGB in 19 Gyr).
The tracks are divided into three groups, 
namely {\it low-mass} stars (${\rm M \le M_{HeF}}$), 
{\it intermediate-mass} stars (${\rm M_{HeF} < M \le M_{up}}$)  and 
{\it high-mass} stars (${\rm M > M_{up}}$). 
The subdivision is made according to the
critical mass at which the ignition of the central fuel
(either helium or carbon) starts quietly depending
on the level of core electron-degeneracy. 
In the adopted tracks, the value of ${\rm M_{HeF}}$ 
depends  slightly on metallicity,
being equal to 1.7, 1.8 and 1.9 ${\rm M_{\odot}}$ for 
Z$=$0.0004, Z$=$0.004 and Z$=$0.008, respectively.
The value of  ${\rm M_{up}}$ is between 5 and 6 ${\rm M_{\odot}}$.
For low-mass stars, 
we have displayed in Figs.~\ref{cmdtracks} and 
\ref{rgbtracks} only the phases 
up to the TRGB in order to avoid excessive confusion.

At the distance of NGC~4449, the MS is sampled only 
for stars with ${\rm M \ga 3 M_{\odot}}$, and corresponds to the 
vertical lines whose 
${\rm m_{F435W}- m_{F555W}}$, ${\rm m_{F555W}- m_{F814W}}$
colors go from $\sim$ $-$0.3 to $\sim$  0 from the 40 ${\rm M_{\odot}}$
to the 3 ${\rm M_{\odot}}$ track. The turnoff is recognizable as a small
blue hook on each evolutionary track of Fig.~\ref{cmdtracks}.
The almost horizontal blue loops correspond to the later
core helium-burning phase for intermediate- and high- 
mass stars, while the bright vertical lines 
in the red  portion of the CMD describe 
the AGB phase.
For low-mass stars, our data sample only the brighter red sequences 
of the RGB, all terminating at the TRGB with approximately 
the same luminosity.
The models considered here do not include the
horizontal feature of the carbon stars in the TP-AGB phase,
recognizable in the ${\rm m_{F814W},  m_{F555W}- m_{F814W}}$
CMD at ${\rm m_{F555W}- m_{F814W} \ga 2}$,
but see \citet{marigo07} for new appropriate models.

The comparison of the CMDs with the tracks 
in Fig.~\ref{cmdtracks} shows that 
the blue plume is populated by high- and intermediate-mass stars on the 
MS, and high-mass stars at the hot edge of the core helium-burning phase.
The red plume samples bright red SGs and AGB stars. The faint 
red sequence at ${\rm m_{F814W} \ga 24}$,
featured in Fig.~\ref{rgbtracks},
is due to low-mass stars in the RGB phase.

The presence of high-, intermediate- and low-mass stars in the CMD
testifies that young, intermediate-age and old stars (several Gyrs) are
present at the same time in NGC~4449.
In particular, the fact that we sample
stars as massive as 40 ${\rm M_{\odot}}$  
implies that the SF was active 5 Myr ago. 
Furthermore, the absence of significant gaps
in the CMD suggests that the SF has been mostly continuous over the
last 1 Gyr. The time resolution gets significantly poorer for higher look-back times 
because of both the intrinsic degeneracy of the tracks
and the large photometric errors at faint magnitudes; thus small 
interruptions in the SF can be easily hidden in the CMD for ages older than
$\ga 1$ Gyr.

The metallicity ${\rm [M/H] =-0.52}$ derived in Section~1 for NGC~4449
corresponds to a metal fraction Z$=$0.005, adopting
Z$_{\odot}=$ 0.017 \citep{sun98}.
However this value 
refers to abundance determinations
in HII regions, thus it is likely to reflect the
metallicity of the youngest stars.
Fig.~\ref{cmdtracks} and ~\ref{rgbtracks} show that
the Z$=$0.004 tracks are in good agreement with
all the phases of the empirical CMDs.
The Z$=$0.0004 and Z$=$0.001
tracks are too blue to account for the observed
RGB feature. 
So, if there is no significant extinction 
intrinsic to NGC~4449, then 
stars with such low metallicities
do not account for a significant 
fraction of the stellar population
in NGC~4449.
Allowing for an age spread from 1 Gyr
to a Hubble time, the Z$=$0.004
tracks are in good agreement with both the 
color and the width of the observed RGB.
\citet{hill98} derived an internal reddening ${\rm E(B-V)
\approx}$ 0.18 from the $H\alpha$/H$\beta$ 
ratio measured in NGC~4449 HII regions.
Adopting this value, the RGB is 
consistent with metallicities as low as 
Z$=$0.001.
However, this extinction value is appropriate for the young star forming
regions, and we expect a lower extinction 
in the regions where 
older stars reside (as demonstrated for the LMC by \citet{zari}).
Either way, the 
Z$=$0.008 models seem slightly too red,
especially in the AGB phase.
Although there are some
uncertainties in the models of this phase, we do believe that this
discrepancy indicates that metallicities higher than Z$=$0.008 
are ruled out for this galaxy.

By superimposing the stellar tracks on the empirical CMDs,
we can attempt a selection of the photometrized stars according to their age.
In Fig.~\ref{cmdsel} we show the four regions selected on the 
${\rm m_{F814W},  m_{F555W}- m_{F814W}}$ CMD
that correspond to different stellar masses,
namely M${\rm \ge 20 \ M_{\odot}}$, 
${\rm 5 M_{\odot} \le M < 20 M_{\odot}}$,
${\rm 1.8 M_{\odot} < M < 5 M_{\odot}}$,
and ${\rm M \le 1.8 M_{\odot}}$.
These regions roughly define the loci of 
{\it very young} stars,
with ages $\la$ 10 Myr, {\it young} stars, with 
${\rm 10 \ Myr \la age \la 100 \ Myr}$, {\it intermediate-age} stars,
with ages between $\sim$ 100 Myr and 1 Gyr, and
{\it old} stars, with ages  $>$ 1 Gyr.
The spatial distribution of the four groups of stars
is shown in Fig.~\ref{spatial}.
We notice that old stars are homogeneously 
distributed over the galaxy,
except for the central regions, where 
the high crowding level makes their detection more difficult.
As we approach younger ages, the distribution 
becomes more and more concentrated.
Stars with ages between 10 and 100 Myr
are highly clustered in an {\it S shaped}
structure centered on the galaxy nucleus.
This could be a bar, which is 
a common feature among Magellanic irregular galaxies.
Stars with ages between 10 and 100 Myr 
also clearly outline the symmetric
structure around the resolved cluster-like object 
(see Fig.~\ref{imagegrid}) that
was already discussed in Section~4. The same structure is visible to a
lesser extent in the spatial distribution of stars with ages 100 Myr
-- 1 Gyr. The resolved cluster-like object itself has a clear RGB (not
shown here), and it is therefore not a young super star cluster.
Very young stars, with ages younger than 10 Myr,
are strongly clustered and detected only in the
very central regions of NGC~4449.
Such young stars should be able to ionize the 
surrounding interstellar
medium and produce HII regions.
Indeed, Fig.~\ref{halpha}, where we plotted the F658N 
(H$\alpha$) image together with the positions
of the stars younger than 10 Myr, shows that
this is the case, with a 
strong correlation between the position of the stars younger than 
10 Myr and the HII  regions.
This indicates that most of the emission is due to
photoionization rather than to shocks, due to, e.g.,
supernovae explosions.

\section{Conclusions}

We have acquired HST/ACS imaging 
in the F435W (B), F555W (V), F814W (I) 
and F658N (${\rm H\alpha}$) filters of the 
Magellanic starburst galaxy NGC~4449
in order to infer its star formation history
and understand the properties of the observed global starburst.
In this paper we present the B, V  and I 
photometry of the resolved stars. 
We detect 299,014 objects in the 
(B,V) CMD, 402,045 objects in the (V,I) CMD,
and 213,099 objects with a measured magnitude
in all the three bands. 
The derived CMDs span a 
magnitude range of $\approx$ 10 mag, 
and sample both the young and the old resolved stellar
population in NGC~4449.
We also detected several candidate clusters 
(at least $\approx$ 40, some of which 
look like very massive globular clusters)
and background galaxies in our images.

We derived a new distance from the TRGB method.
The TRGB is detected at a Johnson-Cousins I
magnitude of ${\rm I_{\rm TRGB} = 24.00 \pm 0.04}$.
At the metallicity of NGC~4449, the TRGB 
is expected at an absolute magnitude of
${\rm M_{\rm I, TRGB} =-3.91}$,
with a systematic error of $\sim 0.15$ mag. This provides 
a distance modulus of ${\rm (m-M)_0 = 27.91 \pm 0.15}$,
i.e. a distance of $3.82 \pm 0.27$ Mpc.
We used also the alternative method of the carbon--star 
luminosity, and found a distance of 
$4.11 \pm 0.38$ Mpc, which is consistent with 
the result from the TRGB method.
Our distance determinations are consistent within the errors
with the value of D$=4.2 \pm 0.5$ Mpc previously provided 
by Karachentsev et al.~(2003).

In the CMDs of NGC~4449 we observe 
a well defined blue plume
(MS and post-MS stars) and red plume 
(red SGs and AGB stars), the horizontal 
tail of the carbon stars in the TP-AGB phase
(in the I, V$-$I CMD), and a prominent RGB.
The presence of all these evolutionary features implies
ages up to at least 1 Gyr and possibly as old as a 
Hubble time.
The comparison of the observed CMDs 
with the Padua stellar evolutionary tracks, 
corrected for the derived distance modulus 
and foreground extinction,
shows that stars as massive as 40 ${\rm M_{\odot}}$  
are present in NGC~4449. Such high masses imply that
the star formation was active 5 Myr ago, and possibly
it is still ongoing.
The absence of significant gaps
in the CMDs suggests also that the star formation
has been mostly continuous over the
last 1 Gyr. However, interruptions in the star formation 
can be easily hidden in 
the CMD for ages older than $\ga 1$ Gyr, because of 
the intrinsic degeneracy of the tracks
and the large photometric errors at faint magnitudes.
The presence of a prominent RGB testifies
that NGC~4449 hosts a population possibly as old as 
several Gyrs or more. However, the color-age degeneracy of the
tracks with increasing look-back time
(at a given metallicity, large age differences
correspond to small color variation in the RGB), and the well
known age-metallicity degeneracy, prevent us
from establishing an exact age for the galaxy.
We will derive a better age estimate  
with the synthetic CMD method, when we'll study the detailed SFH 
(Annibali et al. 2008, in preparation).

Abundance estimates in NGC~4449 HII regions
provide 12 + log(O/H)$ \approx 8.31$
\citep{talent,mar97}, which
corresponds to [O/H] $= -0.52$ if we assume
a solar abundance of 12 + log(O/H)$_{\odot} =$ 8.83 \citep{sun98}. 
These measures are biased toward regions where the 
interstellar medium has been significantly reprocessed, 
and thus are likely to reflect the
metallicity of the youngest generation of stars.
We expect lower metallicity for the oldest
stars in NGC~4449.
Interestingly, though, the Z$=$0.004 stellar evolutionary
tracks (the closest in the Padua set to the metallicity
of NGC~4449), corresponding to ${\rm \log(Z/Z_{\odot}) = -0.63}$
if we adopt ${\rm Z_{\odot}=0.017}$ \citep{sun98},
seem to be in very good agreement
with all the features observed in the empirical CMDs,
 if we assume that there is not significant extinction intrinsic to NGC~4449.
In particular, the next lower metallicity tracks
(at ${\rm Z=0.001}$) are definitively too
blue to account for the observed RGB colors, 
implying that the bulk of the stellar population
older than 1 Gyr was already 
enriched in metals. 

We investigated the presence of age and metallicity 
gradients in NGC~4449. To this purpose,
we divided the total galaxy's field of view 
into 28 rectangular regions of $\approx$
1 kpc$^2$ area, and derived 
the CMDs for the different regions. 
The CMD morphology 
presents a significant spatial dependence:
while the RGB is detected
over the whole field of view of the galaxy,
the blue plume, red plume and blue-loop stars are
present only in the more central regions,
indicating that the stellar population is younger in
the center than in the periphery of NGC~4449.

We also studied the spatial behavior of the carbon--star
luminosity and of the RGB color.
Once the effect of incompleteness and photometric errors
is taken into account, the average magnitude of the C stars 
turns out to be constant within the errors ($\approx$ 0.05 mags). 
This gives an upper limit of $\approx$  0.2 dex
on the metallicity variation over the field of view of NGC~4449.
On the other hand, we find that the RGB 
is intrinsically bluer in the center than
in the periphery of the galaxy.
Bluer RGB colors can be due
to younger and/or  more metal poor stellar populations.
However, as spectroscopic-based 
abundance determinations in galaxies
show that metallicity tends to decrease from the center
outwards, or to remain constant, we 
interpret this as the result 
of a younger, and not more metal poor,
stellar population in the center of NGC~4449.

With the help of the Padua tracks,
we identified in the observed 
${\rm m_{F814W},  m_{F555W}- m_{F814W}}$ CMD
four different zones corresponding
to different stellar masses,
namely M${\rm \ge 20 M_{\odot}}$, 
${\rm 5 M_{\odot} \le M < 20 M_{\odot}}$,
${\rm 1.8 M_{\odot} \le M < 5 M_{\odot}}$,
and ${\rm M < 1.8 M_{\odot}}$.
These regions roughly define the loci of 
stars with age $\la$ 10 Myr, 
${\rm 10 \ Myr \la age \la 100 \ Myr}$,
100 Myr $<$ age $<$ 1 Gyr, and 
age $>$ 1 Gyr.
Low-mass old stars are homogeneously 
distributed over the galaxy's field of view,
with the exception of the central regions, where 
the high crowding level makes their detection
more difficult.
As we approach younger ages, the spatial 
distribution of the stars 
becomes more and more clustered.
Intermediate-age stars (100 Myr $<$ age $<$ 1 Gyr)
are mostly found within $\approx$ 1 kpc from the center.
Stars with ages between 100 Myr and 10 Myr are
found in an {\it S--shaped}
structure centered on the galaxy nucleus,
and extending in the North-South direction
up to 1 kpc away from the center.
 This could be a bar,
which is a common feature among Magellanic 
irregular galaxies.
Stars younger than 10 Myr are very rare, and 
found only in the galaxy nucleus and in the North 
arm of the {\it S shape} structure.
The comparison with the H$\alpha$ image 
shows a tight correlation
between the position of the stars younger than 10 Myr
and the HII regions, indicating that we have identified
the very massive and luminous stars
that ionize the surrounding interstellar
medium. 

\begin{figure}
\epsscale{1}
\plotone{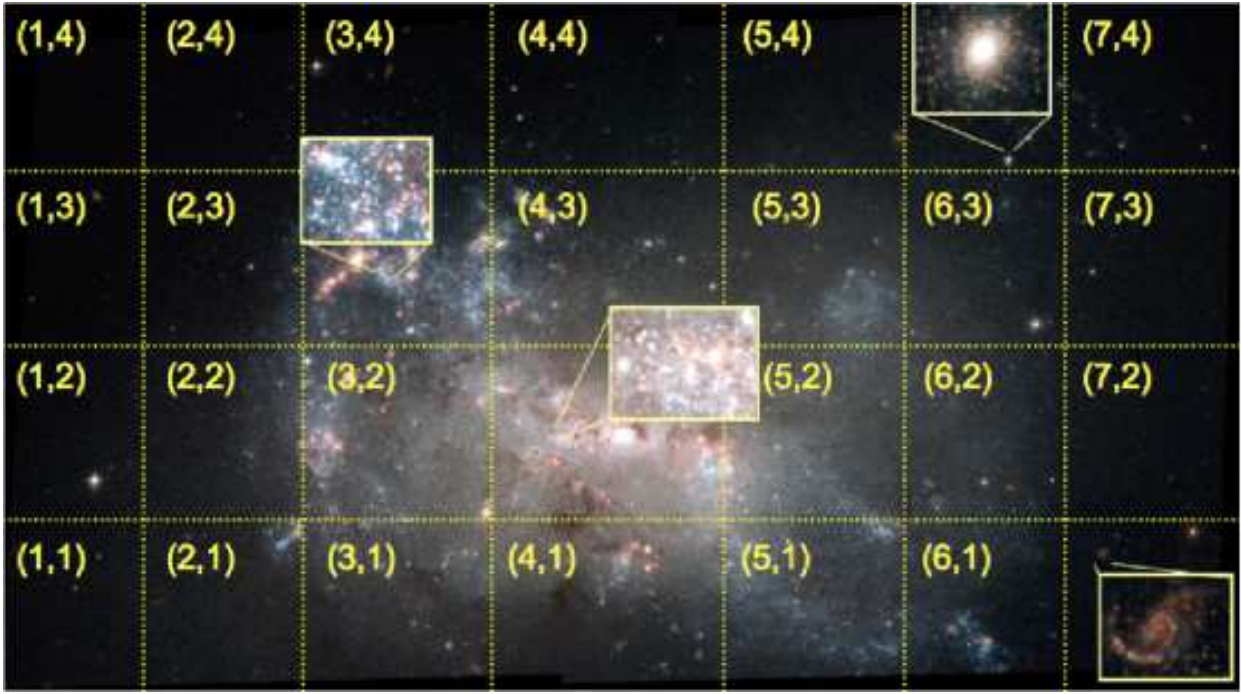}
\caption{Mosaicked image of NGC~4449
where the dotted lines indicate the adopted 
subdivision into (4 $\times$ 7) regions.
The insets show blowups of $\sim$ 5 $\times$ 5
${\rm arcsec^2}$ regions to provide an idea of both the amount
of crowding and the resolution of our data.
At the distance of NGC~4449, each region samples
an area of ${\rm \approx 1 \ kpc^2}$.
\label{imagegrid}}
\end{figure}

\begin{figure}
\epsscale{1}
\plotone{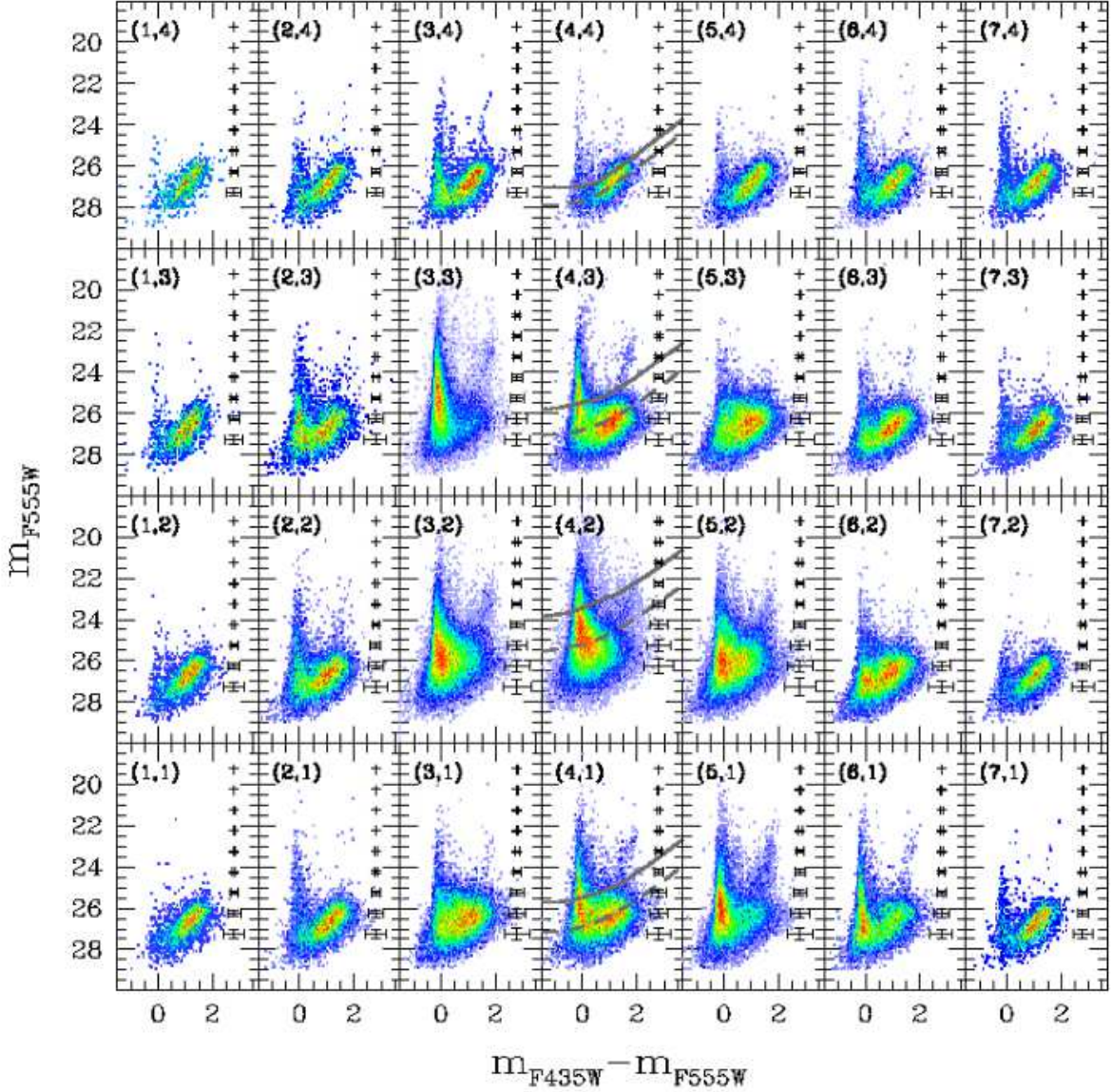}
\caption{ ${\rm m_{F555W}}$ versus ${\rm m_{F435W}- m_{F555W}}$ CMDs
for the 28 regions selected in the field of NGC~4449.
The CMDs are represented through Hess diagrams. 
The number density of stars in the 
${\rm m_{F555W}}$, ${\rm m_{F435W}- m_{F555W}}$
plane increases from blue to red. 
We indicate the average photometric errors
as derived from the artificial star experiments for each region.
The color error is evaluated at  ${\rm m_{F435W}- m_{F555W}=1}$.
The 90 \% (solid line) and 50 \% (dashed line)
completeness levels are indicated
on the CMDs of the central column.
\label{hessregionbv}}
\end{figure}

\begin{figure}
\epsscale{1}
\plotone{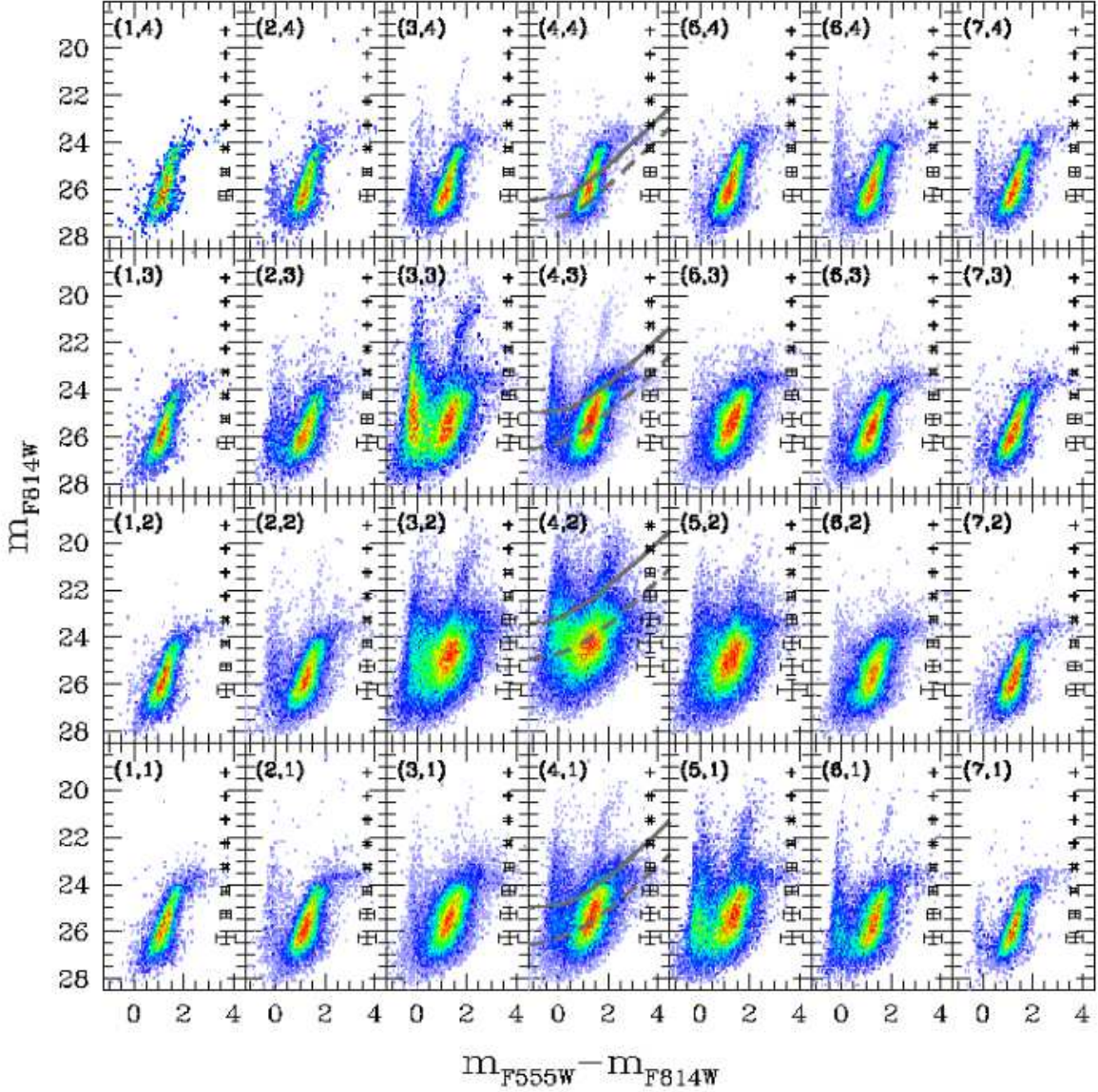}
\caption{ ${\rm m_{F814W}}$ versus ${\rm m_{F555W}- m_{F814W}}$ CMDs
for the 28 regions selected in the field of NGC4449.
The CMDs are represented through Hess diagrams. 
The number density of stars in the 
${\rm m_{F814W}}$, ${\rm m_{F555W}- m_{F814W}}$
plane increases from blue to red. 
We indicate the average photometric errors
as derived from the artificial star experiments for each region.
The color error is evaluated at  ${\rm m_{F555W}- m_{F814W}=1}$.
The 90 \% (solid line) and 50 \% (dashed line)
completeness levels are indicated
on the CMDs of the central column.
\label{hessregionvi}}
\end{figure}

\begin{figure}
\epsscale{1}
\plotone{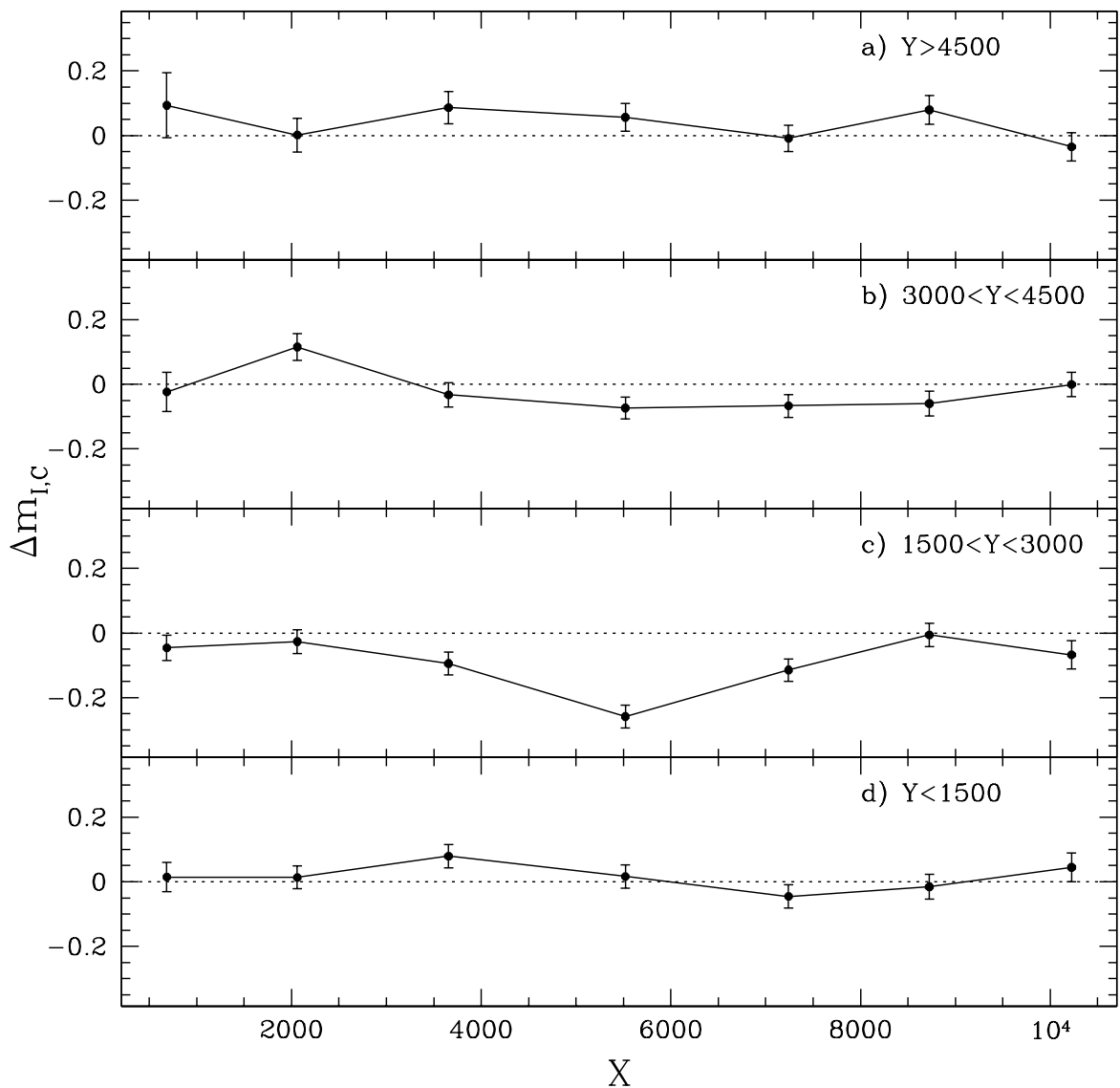}
\caption{ 
Variation of the carbon star luminosity as a function
of position in the field of NGC~4449. The quantity 
$\Delta m_{I, carbon}$ along the ordinate is the difference 
between the carbon star F814W magnitude measured in a specific region, 
and the carbon star magnitude averaged over 
the whole field of view of NGC~4449. 
Along the abscissa is the X coordinate in pixels. 
From top to bottom, the panels refer to regions of 
decreasing Y coordinate. 
\label{Clum}}
\end{figure}

\begin{figure}
\epsscale{0.9}
\plotone{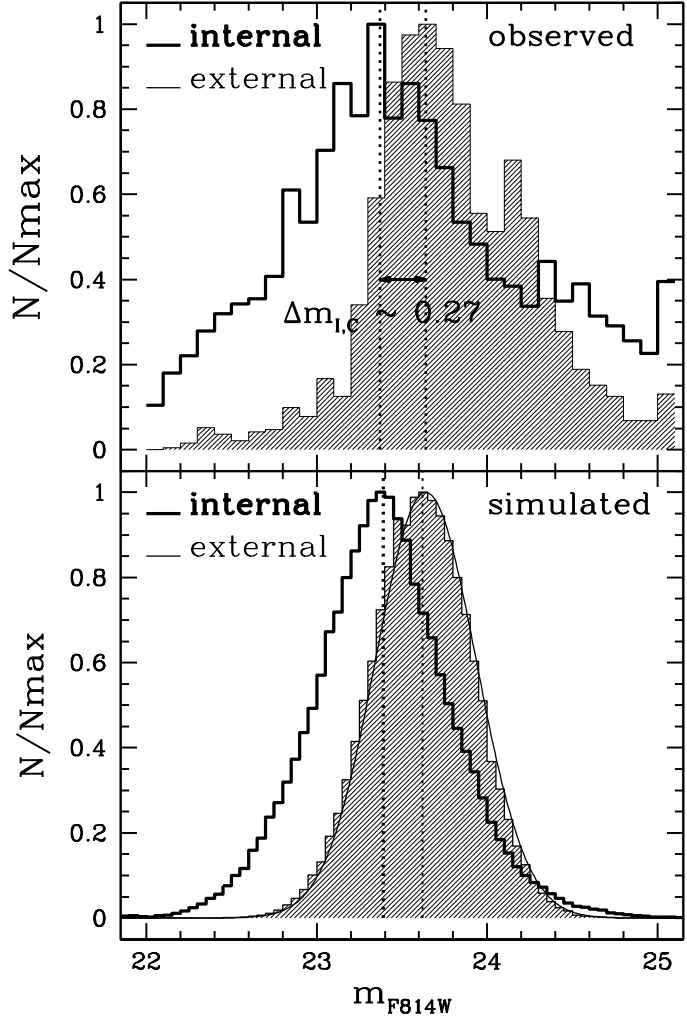}
\caption{  
{\bf Top panel:} 
observed C star luminosity functions (LFs)
for an external region 
((1,1:4), (1:7,4), (7,1:4))
of NGC~4449 (shaded histogram), 
and for the most internal one (4,3).
The peaks of the external and internal 
distributions, as derived from Gaussian fits,
are ${\rm m_{F814}=23.64}$ and ${\rm m_{F814}=23.37}$,
respectively. The external distribution is
narrower (${\rm \sigma_{F814W} \approx 0.3}$) 
than the internal one (${\rm \sigma_{F814W} \approx 0.5}$). 
{\bf Bottom panel:}
simulated distribution for the C stars.
The Gaussian curve is the assumed initial 
${\rm m_{F814}}$ distribution for the C stars
(${\rm m_{F814, 0}=23.64, \sigma_{F814W} = 0.3}$).
The shaded and thick line histograms represent 
the resulting distributions 
in the external and internal regions, after errors 
and incompleteness have been applied.
The peak of the simulated internal distribution has
a shift of ${\rm \Delta m_{F814W} \approx -0.25}$ 
with respect to the peak of the Gaussian curve.
Its width ($\sigma \approx 0.4$) is larger than the
Gaussian, due to the larger photometric errors 
in the center of NGC~4449.
\label{Clf}}
\end{figure}

\begin{figure}
\epsscale{1}
\plotone{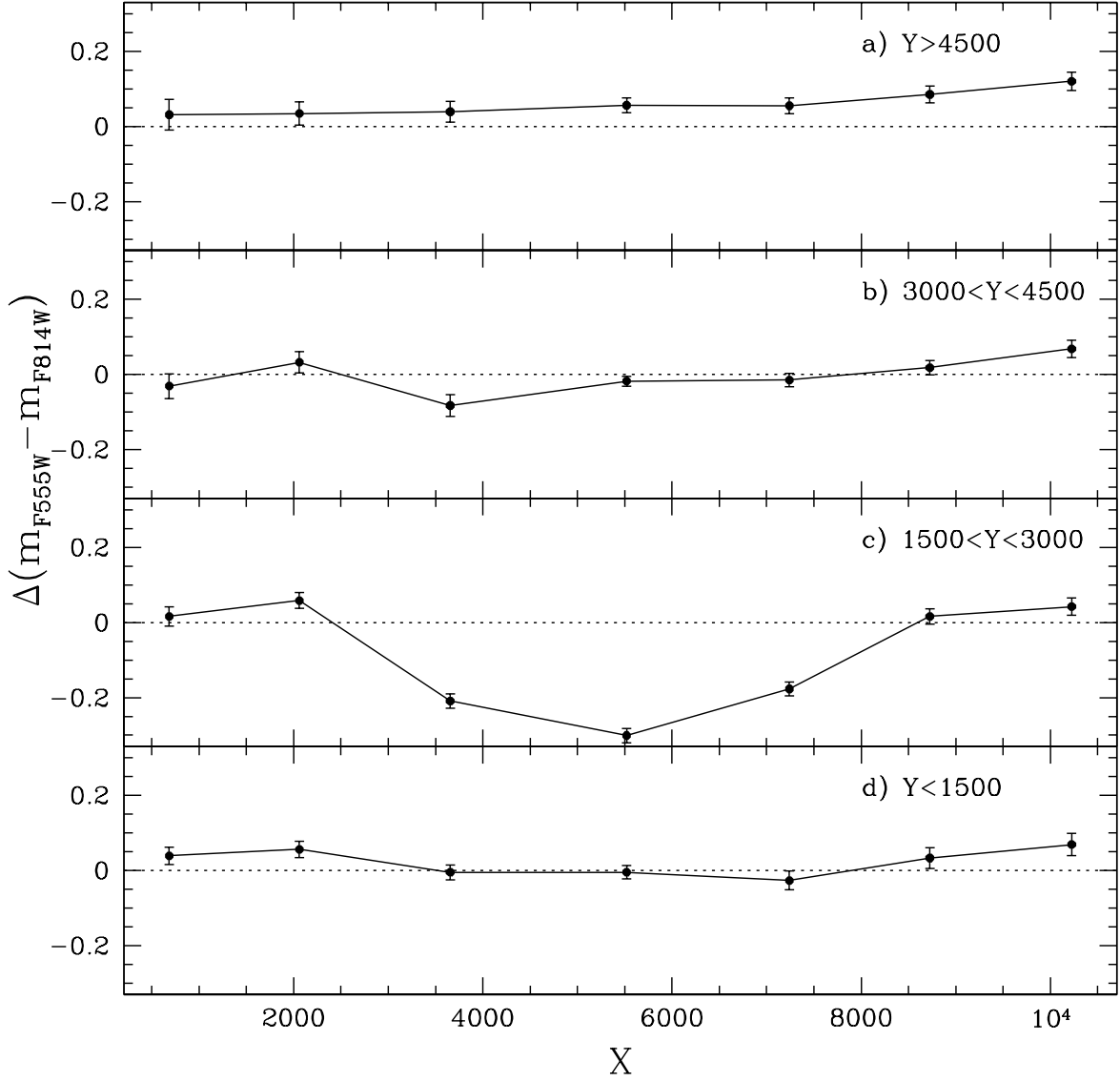}
\caption{${\rm m_{F555W}- m_{F814W}}$ color variation of the RGB 
as a function of position in NGC~4449 field.
The RGB colors were computed performing an average 
between ${\rm 24 \la m_{F814W} \la 25}$, i.e. within one 
magnitude below the TRGB.
The variation ${\rm \Delta m_{F555W}- m_{F814W}}$
was computed with respect to 
the RGB color in the total 
field of view of NGC~4449.
Along the abscissa is the X coordinate in pixels.
From top to bottom, the panels refer to regions of decreasing Y
coordinate.
\label{deltargb}}
\end{figure}

\begin{figure}
\epsscale{0.7}
\plotone{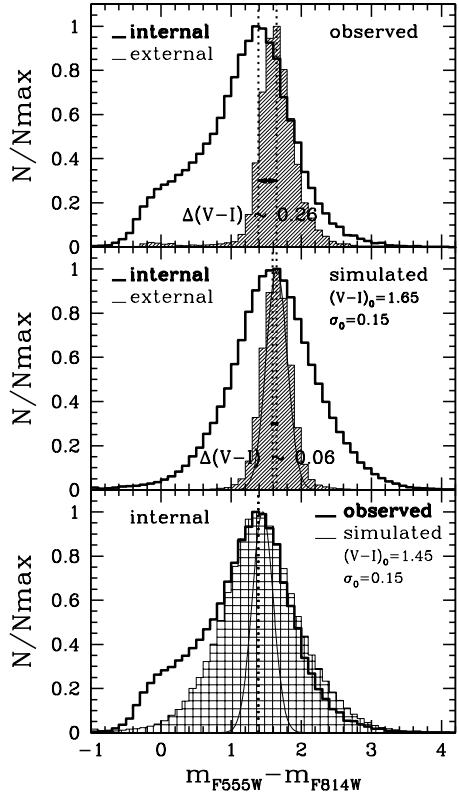}
\caption{  
{\bf Top panel:} Observed 
color distributions of stars 
with ${\rm 24 \le m_{F814} \le 25}$ mag,
for an external region 
((1,1:4), (1:7,4), (7,1:4), shaded histogram), 
and for an internal region 
((3:5,2) thick line) of NGC~4449.
The distributions, which peak at red 
(${\rm m_{F555W} - m_{F814W} > 1}$) colors,
denote the presence of old low-mass stars
in the RGB phase. The dotted lines indicate
the position of the peaks 
(as derived from a Gaussian fit) of the
external and internal distributions,
at ${\rm m_{F555W} - m_{F814}=1.65}$ and 
${\rm m_{F555W} - m_{F814}=1.39}$,
respectively . 
{\bf Central panel:}
simulated color distributions.
The Gaussian curve is an assumed intrinsic 
distribution, with
peak and standard deviation
as indicated in the panel.
The shaded and thick line histograms are
the simulated distributions 
in the external and internal regions, 
respectively. They were generated 
through Monte Carlo extractions from the
assumed Gaussian,
and applying photometric errors and 
completeness levels as derived 
from artificial star experiments
in the considered regions.
The dotted lines indicate the peaks
of the simulated distributions.
While the peak of the external distribution
is the same as that of the initial
Gaussian, the peak of the internal
distribution is blueshifted by
$\approx$ 0.06 mags.
{\bf Bottom panel:} 
The thick line is the observed color
distribution for the internal region 
(3:5,2). The shaded histogram is a new
simulated distribution for the central region
drawn from a Gaussian
with a bluer peak.
The peaks of the observed and simulated
distributions, as indicated 
by the dotted line,
are now the same.
\label{rgbcolor}
}
\end{figure}

\begin{figure}
\epsscale{0.6}
\plotone{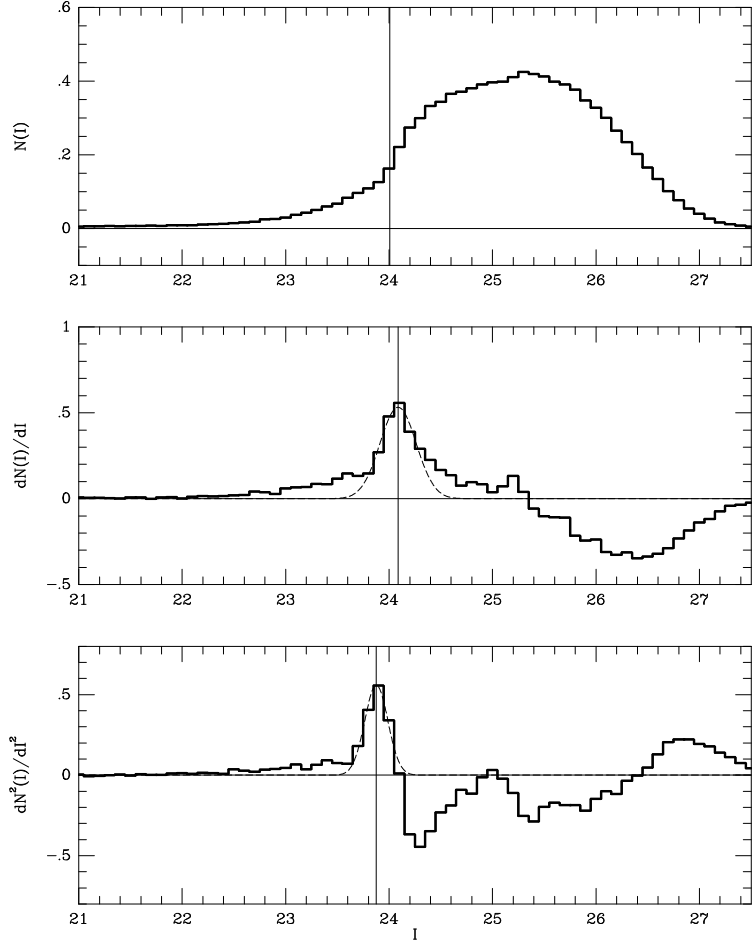}
\caption{Top panel: extinction corrected $I$-band luminosity
function (LF) of stars with V$-$I in the range $1.0$--$2.0$
(solid). The solid line is our final estimate of the TRGB magnitude,
obtained as described in the text. 
Middle panel: first-order derivative of the
LF (solid), with the best-fitting Gaussian overplotted (dashed). The
vertical line indicates the center of the Gaussian, which provides a
slightly biased overestimate of the TRGB magnitude. 
Bottom panel: second-order
derivative of the LF (solid), with the best-fitting Gaussian
overplotted (dashed). The vertical line indicates the center of the
Gaussian, which provides a slightly biased underestimate of the TRGB
magnitude. The vertical scales in all panels are normalized
arbitrarily.
\label{trgb}}
\end{figure}

\begin{figure}
\epsscale{1}
\plotone{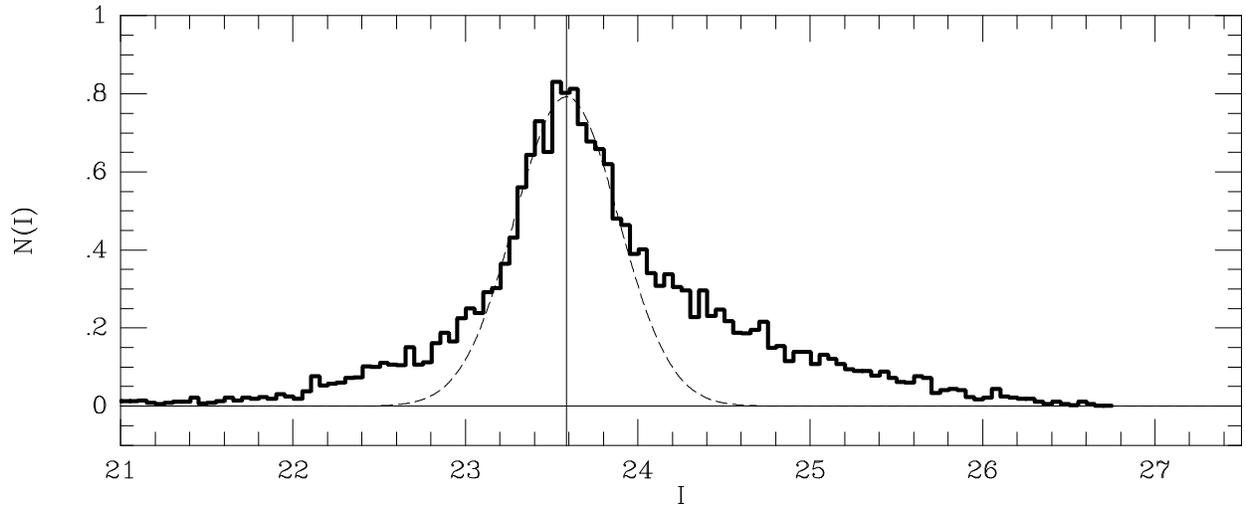}
\caption{Extinction corrected $I$-band luminosity function
(LF) of stars with V$-$I in the range $2.2$--$3.0$ (solid), with the
best-fitting Gaussian overplotted (dashed). The vertical line
indicates the center of the Gaussian, which is an estimate of the
average $I$-band magnitude of the carbon stars in NGC 4449. The
vertical scale is normalized arbitrarily.
\label{agbdist}}
\end{figure}

\begin{figure}
\epsscale{0.8}
\plotone{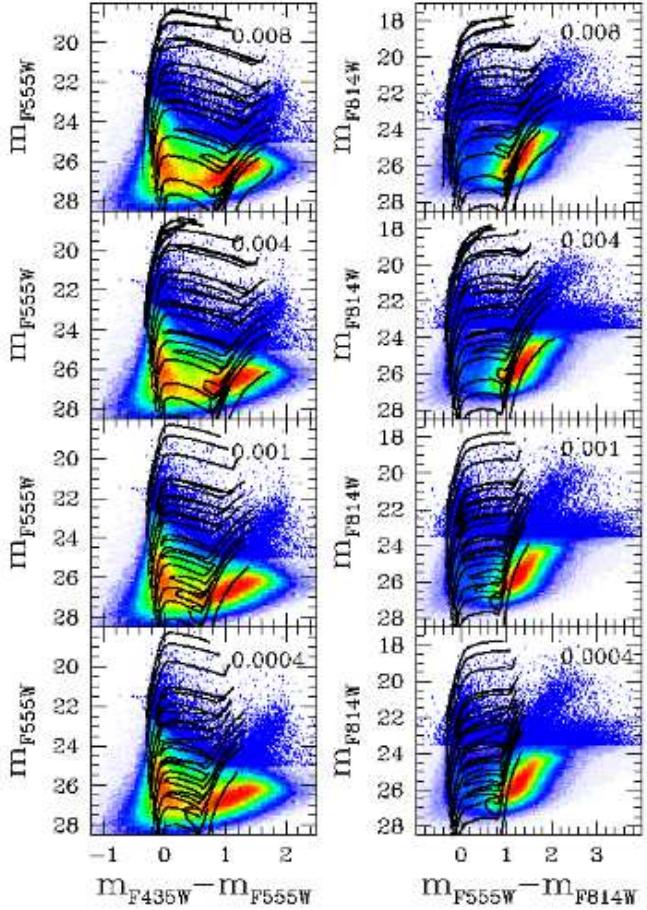}
\caption{${\rm m_{F555W}}$, ${\rm m_{F435W}- m_{F555W}}$ CMDs
(left panels) and 
${\rm m_{F814W}}$, ${\rm m_{F555W}- m_{F814W}}$ CMDs 
(right panels) of NGC~4449, with superimposed
stellar evolutionary tracks for different 
metallicities, as indicated in the panels.
The CMDs are represented through Hess diagrams,
with the number density of stars 
increasing from blue to red. 
Points have been used to plot the brightest stars.
The Z$=$0.008, Z$=$0.004 and Z$=$0.0004 models
are the Padua stellar evolutionary tracks 
(Fagotto et al. 1994a, 1994b)
converted into the ACS filters
by applying bolometric corrections, 
and adopting a foreground ${\rm E(B-V)=0.019}$ and 
distance modulus ${\rm (m-M)_0 = 27.91}$.
The Z$=$0.001 tracks were obtained from the Padua tracks
through linear interpolation in metallicity \citep{ang06}.
We displayed the following stellar masses:
(left to right): 40, 30, 20, 12, 9, 7, 5, 4, 3, 2 and 1  $M_{\odot}$.
The corresponding lifetimes (which  slightly depend on metallicity)
are: 5, 7, 10, 20, 35, 56, 110, 180, 370, 1120 and 8410 Myr, respectively.
No extinction was assumed intrinsic to NGC~4449.
\label{cmdtracks}}
\end{figure}

\begin{figure}
\epsscale{0.8}
\plotone{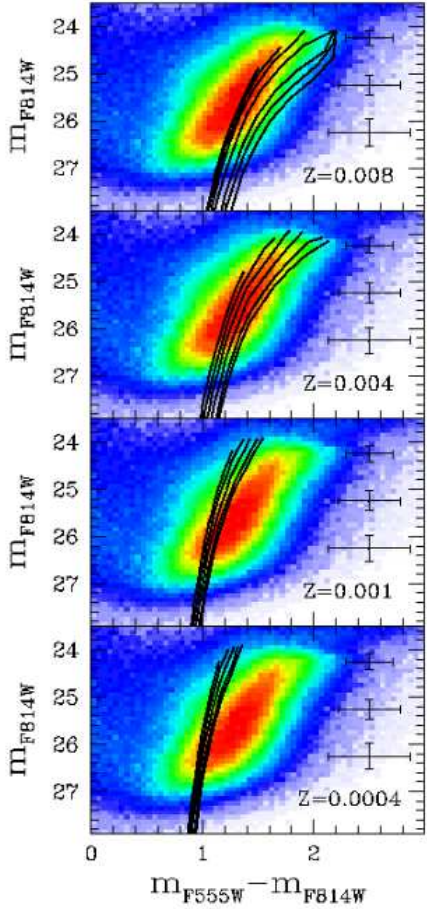}
\caption{  
${\rm m_{F814W}}$, ${\rm m_{F555W}- m_{F814W}}$ CMD
of NGC~4449 showing the RGB feature, 
with superimposed the Padua 
low-mass stellar evolutionary tracks 
for different 
metallicities, as indicated in the panels.
The CMDs are represented through Hess diagrams,
with the number density of stars 
increasing from blue to red. 
The foreground extinction and the distance modulus
adopted to display the tracks are
${\rm E(B-V)=0.019}$ and ${\rm (m-M)_0 = 27.91}$.
We plot the following stellar masses:
(left to right): 
1.8, 1.6, 1.4, 1.2, 1 and  0.9 $M_{\odot}$.
The corresponding lifetimes (which  slightly depend on metallicity)
are: 1300, 1630, 2520, 4260, 8410, 12500 Myr, respectively.
No extinction was assumed intrinsic to NGC~4449.
\label{rgbtracks}}
\end{figure}

\begin{figure}
\epsscale{1}
\plotone{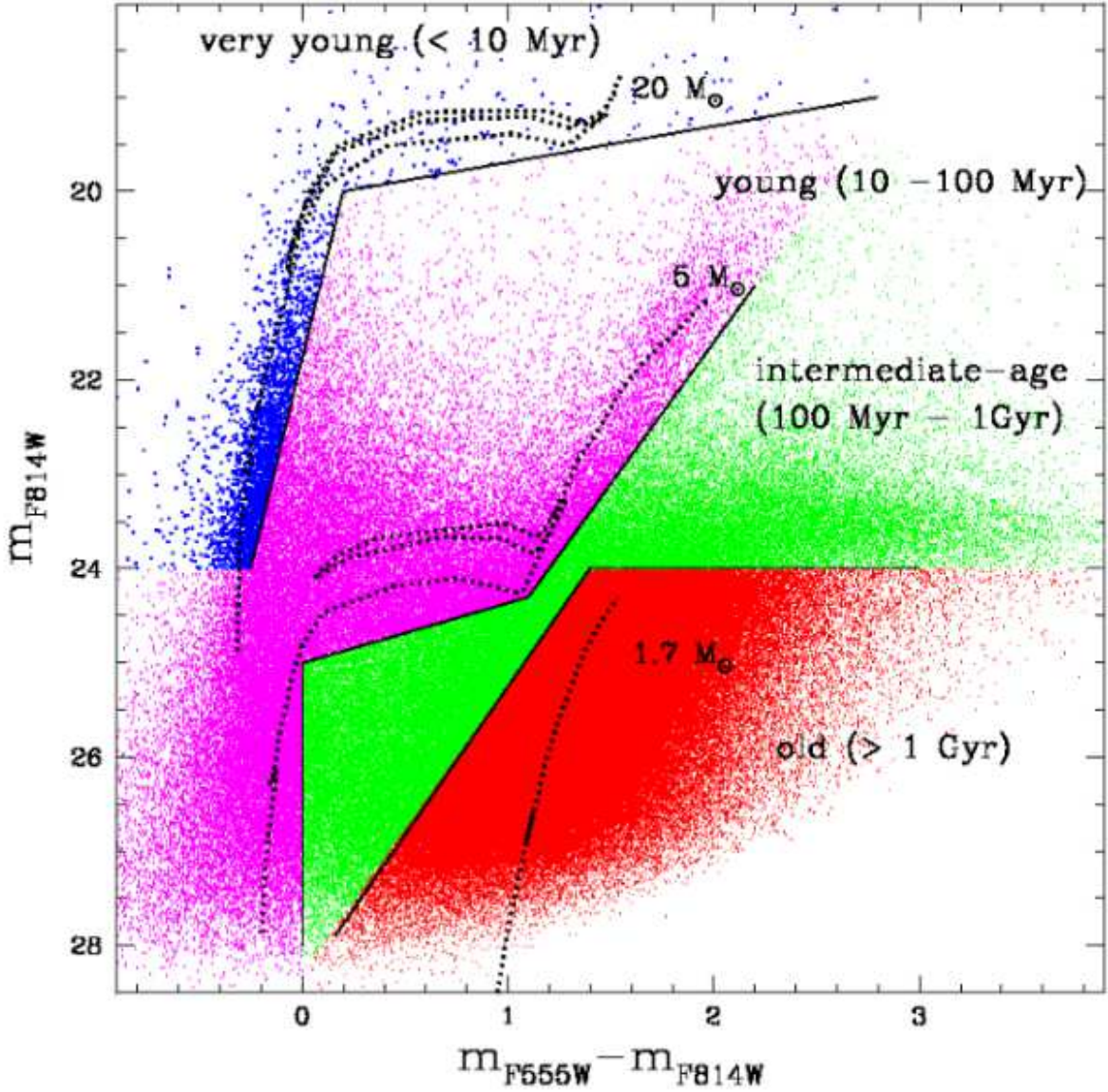}
\caption{${\rm m_{F814W}}$, ${\rm m_{F555W}- m_{F814W}}$ CMD
of NGC~4449 (402,045 stars), where we have identified 
the loci approximately corresponding
to very young stars (age $\la$ 10 Myr),
young stars (10 Myr $\la$ age $\la$ 100 Myr),
intermediate-age stars
(100  Myr $\la$ age $\la$ 1 Gyr), and old stars
(age $\ga$ 1 Gyr).
On the same CMD, we plotted the 
stellar evolutionary tracks for the masses
that indicatively define the age boundaries.
\label{cmdsel}}
\end{figure}

\begin{figure}
\epsscale{1}
\plotone{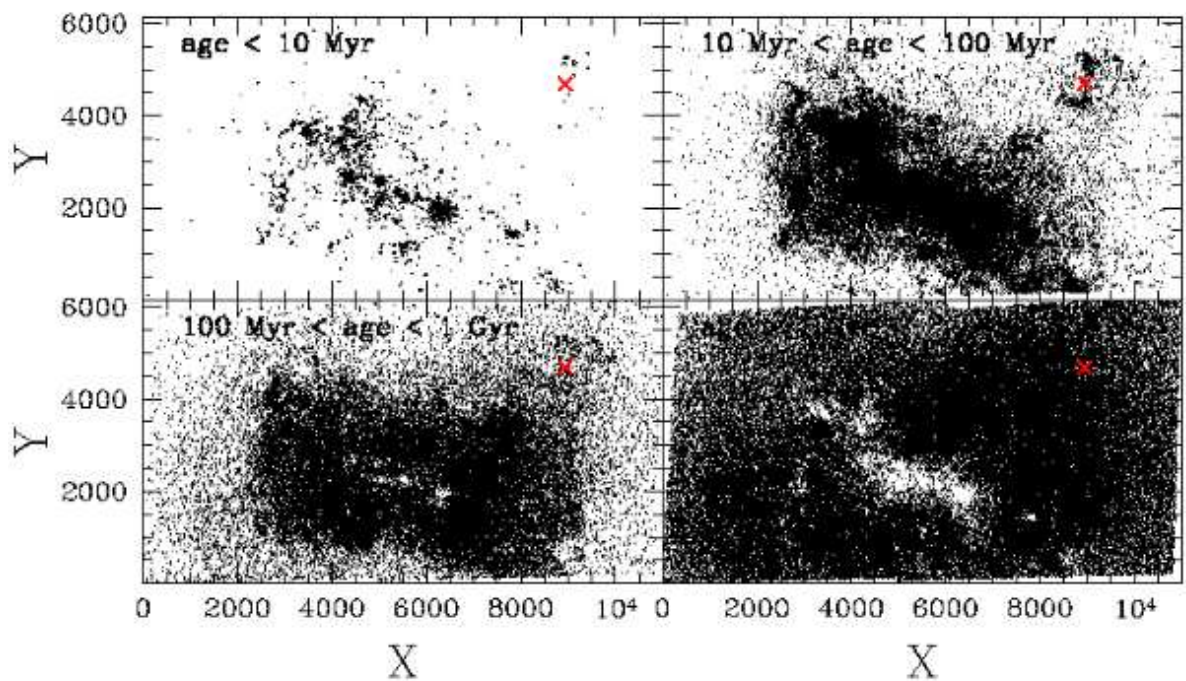}
\caption{Spatial distribution of very young
(age $\la$ 10 Myr), young 
 (10 Myr $\la$ age $\la$ 100 Myr), intermediate-age 
(100  Myr $\la$ age $\la$ 1 Gyr), and old stars
(age $\ga$ 1 Gyr) in NGC~4449. 
Stars were selected according to their position
in the ${\rm m_{F814W}}$, ${\rm m_{F555W}- m_{F814W}}$ CMD,
as shown in Figure \ref{cmdsel}.
Notice that crowding prevents us from detecting old stars
in the central regions.
The red X marks the position of the resolved
cluster-like object for which a blow-up is shown in Figure~\ref{imagegrid}.
This object is surrounded by a bi-symmetric structure that is clearly
outlined by stars with ages from 10--100 Myr, and to a lesser extent
by stars of ages 100 Myr -- 1 Gyr.
\label{spatial}}
\end{figure}

\begin{figure}
\epsscale{1}
\plotone{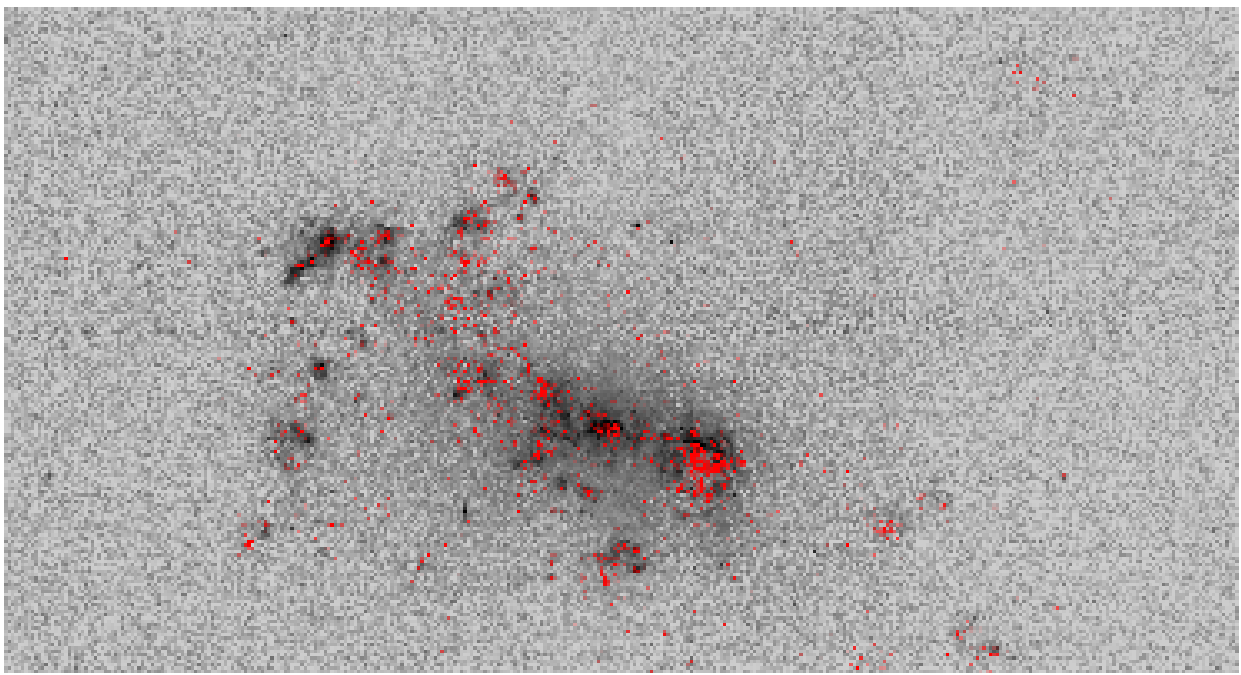}
\caption{ACS-F658N  (H$\alpha$) image of NGC~4449 
plotted as a greyscale map. 
The positions of the stars younger than $\approx$ 10 Myr are
superimposed using red dots.
\label{halpha}}
\end{figure}

One of the many star clusters visible in our image is of
particular interest. This cluster on the West side of the galaxy is
surrounded by a symmetric structure that is particularly well
outlined by stars with ages in the range of 10--100 Myr (see
Fig.~\ref{spatial}). This structure could be due to tidal tails or 
spiral--like feature associated with a dwarf galaxy that is currently being
disrupted by NGC~4449. The cluster could be the remnant nucleus of
this galaxy. It is resolved into red stars, and it has a significant
ellipticity (see Fig.~\ref{imagegrid}). This is reminiscent of the star cluster
$\omega$ Cen in our own Milky Way, which has also been suggested to be
the remnant nucleus of a disrupted dwarf galaxy \citep{omegacen}.
More details about this object will be presented in a forthcoming
paper (Aloisi et al. 2008, in preparation).

Quantitative information on the star formation history of NGC~4449
is fundamental in order to understand the connection
between the global starburst observed and processes
such as merging, accretion and interaction.
The star formation history of NGC~4449 will be derived
in a forthcoming paper 
through the synthetic CMD method, which is based on stellar evolutionary
tracks, and is able to fully account for the effect of 
observational uncertainties, such as photometric errors,
blending and incompleteness of the observations.





\acknowledgments

Support for proposal \#10585 was provided by NASA
through a grant from STScI, which is operated by
AURA, Inc., under NASA contract NAS 5-26555.
We thank Livia Origlia for providing the photometric conversion tables 
to the ACS Vegamag system.





\appendix




\clearpage



\clearpage

\end{document}